\documentclass[journal,article,submit,pdftex,moreauthors]{Definitions/mdpi} 
\firstpage{1} 
\pubvolume{1}
\issuenum{1}
\articlenumber{0}
\pubyear{2024}
\copyrightyear{2024}
\datereceived{ } 
\daterevised{ } 
\dateaccepted{ } 
\datepublished{ } 
\hreflink{https://doi.org/} 
\Title{Quark clusters, QCD vacuum and the cosmological $^7$Li, Dark Matter and Dark Energy problems}

\TitleCitation{Quark clusters, QCD vacuum and the cosmological $^7$Li, Dark Matter and Dark Energy problems}

\Author{Rachid Ouyed $^{1,*}$, Denis Leahy $^{1}$, Nico Koning $^{1}$ and Prashanth Jaikumar $^{2}$}

\AuthorNames{Rachid Ouyed, Denis Leahy, Nico Koning and Prashanth Jaikumar}

\AuthorCitation{Ouyed, R.; Leahy, D.; Koning, N.; Jaikumar, P.}

\address{%
$^{1}$ \quad Department of Physics and Astronomy, University of Calgary, 2500 University Drive NW, Calgary, AB T2N 1N4, Canada; leahy@ucalgary.ca (D.L.); nakoning@ucalgary.ca (N.K.)\\
$^{2}$ \quad Department of Physics and Astronomy, California State University Long Beach, 1250 Bellflower Blvd., Long Beach, CA 90840, USA; prashanth.jaikumar@csulb.edu (P.J.)}

\corres{Correspondence: rouyed@ucalgary.ca}

\abstract{We propose a non-exotic electromagnetic solution (within the standard model of particle physics) to the cosmological $^7$Li problem based upon  a narrow 2 MeV photo-emission line  from the decay of light Glueballs (LGBs). These LGBs form within color superconducting, tens of Fermi in size,  quark clusters (SQCs; with baryon number $A_{\rm sqc}\sim 10^6$) in the  radiation-dominated post-BBN epoch. The mono-chromatic line from the $LGB\rightarrow \gamma+\gamma$ decay reduces   Big-Bang nucleosynthesis (BBN) $^7$Be by $2/3$ without affecting other abundances or Cosmic Microwave Background (CMB) physics, provided the combined mass of the SQCs is greater than the total baryonic mass in the Universe. 
  Following the LGB emission, the in-SQC Quantum-ChromoDynamics (QCD) vacuum becomes unstable and ``leaks" (via quantum tunnelling) into
   the external space-time (trivial)  vacuum inducing a decoupling of SQCs from hadrons.  In seeking a solution to the $^7$Li problem, we uncovered a solution which also addresses the dark energy (DE) and dark matter (DM) problem making these critical problems intertwined in our model.  Being colorless, charge neutral, optically thin and transparent to hadrons, SQCs interact only gravitationally making them a viable cold DM (CDM) candidate.
       The leakage (i.e. quantum tunnelling) of the in-SQC  QCD vacuum to the trivial vacuum offers an explanation of DE in our model and allows for a cosmology which evolves into a $\Lambda$CDM universe at low redshift with a possible resolution of the Hubble tension.     Our model distinguishes itself by proposing that the QCD vacuum within SQCs possesses the ability to tunnel into the exterior trivial vacuum, resulting in the generation of DE. This  implies the possibility that DM and hadrons might represent distinct phases of quark matter within the framework of QCD, characterized by different vacuum properties. We discuss SQC formation in heavy-ion collision experiments  at moderate temperatures and the possibility of detection of  MeV photons from the $LGB\rightarrow \gamma+\gamma$ decay.}


\begin{document}

\section{Introduction}
\label{sec:introduction}

The primordial abundances of the light elements produced in the first few minutes 
of the universe predicted by standard hot Big-Bang cosmology \cite{hoyle_1964,peebles_1966,wagoner_1967} are in excellent 
agreement with the abundances inferred from data (e.g. \cite{tytler_2000}).
 Big Bang Nucleosynthesis (BBN) starts when the deuteron (D) bottleneck is overcome at\footnote{Dimensionless quantities are defined as $f_x = f/10^x$ with quantities in cgs units unless specified otherwise.} $k_{\rm B}T\sim 100$ keV and terminates at $k_{\rm B}T \sim 30$ keV (redshift $z\sim 4\times 10^8$)  due to electrostatic repulsion between nuclei (e.g. \cite{lang_1999}); $k_{\rm B}$ is the Boltzmann constant. Significant amounts of D, $^3$H and $^4$He build up followed by the production of much less-abundant elements
   such as $^7$Be. With a  half-life of $\sim$ 53 days, $^7$Be decays into $^7$Li via bound electron capture, with emission of a neutrino (e.g. \cite{khatri_2011}).   This cannot occur, however, until recombination at  $z\sim 1100$ when $^7$Be becomes singly ionized. 
 The measured $^7$Li abundance is $\sim$ 1/3 of what is expected from this process \cite{spite_1982}
  defining  the cosmological $^7$Li problem  (see \cite{fields_2011} for a review). 
 
 The standard theory of electromagnetic cascades onto a photon background predicts a quasi-universal shape for the resulting non-thermal photon spectrum (e.g. \cite{berezinsky_1990}).  In the case of non-thermal BBN,  cosmological constraints using this quasi-universal shape make purely electromagnetic solutions to the $^7$Li problem impossible (e.g. \cite{kawasaki_1995}) unless the injected photon energy falls below the pair-production threshold in which case the spectral shape is very different (see \cite{poulin_2015} and references therein).  The effective pair production threshold is
  $E_{\gamma, {\rm pair}}\simeq (m_ec^2)^2/22k_{\rm B}T \sim 12\ {\rm MeV}\times ({\rm keV}/T)$ below which the double photon pair-creation process receives a Boltzmann suppression (\cite{kawasaki_1995}; see also \cite{protheroe_1995}); $m_{\rm e}$ is the electron mass.  
If injected when BBN is over,  sub-threshold  photons act to post-process the abundances
computed in the standard scenario. Photons injected with  $1.59\ {\rm MeV} < E_{\gamma} < 2.22$ MeV can destroy $^7$Be  (suppressing $^7$Li production) without affecting other BBN abundances, distorting the CMB background or injecting excess entropy \cite{poulin_2015,kawasaki_2020}.

In this paper, we sketch a model for the injection of sub-threshold photons ($E_{\gamma}< E_{\gamma, {\rm pair}}$) in the radiation-dominated post-BBN epoch, characterized by a universe temperature in the keV range. The condition $2\ {\rm MeV} < E_{\gamma, {\rm pair}}$ implies $T_{\rm CMB}< \sim 6$ keV, which corresponds to the post-BBN era when the universe is a few hours old and has a redshift below $6\ {\rm keV}/T_{\rm CMB, 0}\sim 2.5\times 10^7$ where $T_{\rm CMB,0}\sim 2.73$ keV.
 This model focuses on the nearly instantaneous emission (with a timescale of $\sim 10^{-17}$ s) of a mono-chromatic 2 MeV line by quark clusters (QCs). These QCs are formed during the early universe's Quantum-ChromoDynamics (QCD) cross-over phase transition at a redshift of $z_{\rm QCD}\simeq T_{\rm QCD}/T_{\rm CMB, 0}\sim 6.7\times 10^{11}$, when the universe was a few microseconds old and had a temperature of $T_{\rm QCD}\sim 150$ MeV.

Our model is grounded in the fundamental principles of QCD and relies on the properties of quark matter in a regime where quarks are deconfined and form Cooper pairs. Prior to delving into the details of our model, we provide a didactic introduction to the QCD phase diagram, the quark-gluon plasma (QGP), and other pertinent phases of quark matter in Appendix \ref{appendix:pedagogical} (titled "Pedagogical Framework"). Additionally, we offer a general overview of our model and the resulting interconnectedness of the $^7$Li problem, Dark Energy (DE), and Dark Matter (DM) enigmas.

In our study, QCs consist of collections of up and down quarks. To ensure charge neutrality, the number density of down quarks (with a charge of $-1/3$) is twice that of up quarks (with a charge of $+2/3$). These QCs, which are assemblies of the quark-gluon plasma (QGP), have a radius of approximately 100 Fermi, rendering them transparent to photons.

Being composed of quarks, QCs are classified as baryons and are thus encompassed within the standard model of particle physics. They can be envisioned as macroscopic nucleons (not nuclei) carrying baryon number entirely through the quarks. Here and throughout the paper, baryonic matter includes all quarks which is more general than just protons and neutrons.  The density of QCs is on the order of $\sim 10^{39}$ cm$^{-3}$, which is approximately ten times the nuclear saturation density. Consequently, an approximately 100 Fermi QC possesses a baryon number $A_{\rm qc}\sim 10^6$. QCs are distinct entities from the significantly larger ($A>>A_{\rm qc}$) cosmic strange-quark nuggets (\cite{witten_1984}) and Axion quark nuggets (\cite{zhitnitsky_2003}), requiring different formation mechanisms (see bullet point \#\ref{itemFormation} in Section \ref{sec:discussion}).

In the post-BBN era, QCs transition into a color superconducting (CSC) phase, where their constituent quarks form Cooper pairs (see Appendix \ref{appendix:pedagogical})  and give rise to what we refer to as Superconducting Quark Clusters (SQCs). For the purposes of this paper, we select the two-flavor color superconducting (2SC) phase as the reference phase, allowing us to convey the essence of our idea.
 Within the 2SC phase, a significant fraction of gluons (3/8 of them) do not interact with the paired quarks. Instead, they form their own condensate in the form of light glueballs (LGBs). Unlike the confined quark matter where GeV-scale glueballs are anticipated, LGBs within the 2SC phase exhibit a mass ($M_{\rm LGB}$) in the MeV range.

Crucially, in the 2SC phase, it can be demonstrated that LGBs are electromagnetically unstable and decay into photons via the process $LGB\rightarrow \gamma+\gamma$ (by coupling with virtual quark loops which carry the electric charge). With a reasonable choice of QCD parameters, the 2SC phase accommodates LGBs with a mass of approximately $M_{\rm LGB}c^2\sim 4$ MeV.

During the post-BBN epoch, where $kT_{\rm CMB}<<M_{\rm LGB}c^2$, the decay of these MeV-scale LGBs generates a narrow, mono-chromatic $\sim 2$ MeV line. This line emission presents a promising avenue for potentially resolving the enduring $^7$Li problem.

The total mass of SQCs required to address the $^7$Li problem is comparable to the observed value of Dark Matter (DM) in the Universe.  Furthermore, the distinctive properties of SQCs - being colorless, cold, charge-neutral, optically thin, and decoupled from hadrons - position them as a plausible candidate for Cold Dark Matter (CDM).

We propose that the decay of LGBs within SQCs may lead to the destabilization of the in-SQC Quantum QCD vacuum. This destabilization, in turn, causes the in-SQC vacuum to undergo quantum tunnelling into the external (trivial) vacuum of space-time, a phenomenon we refer to as "leakage." Notably, this leakage process involves the transition to a vacuum state characterized by vanishing QCD (quark and gluon) condensates.

The tunnelling of the in-SQC QCD vacuum, behaving as Dark Energy (DE), occurs in the exterior space-time within our model. At low redshifts, the cosmological behavior aligns with the $\Lambda$CDM model. Therefore, in our model, the loss of gluonic content from SQCs due to LGB decay not only provides a potential solution to the $^7$Li problem but also triggers the leakage process, leading to a cosmological scenario encompassing both Cold Dark Matter (CDM) in the form of SQCs and Dark Energy (DE) arising from the tunnelling of the in-SQC QCD vacuum.

The paper is structured as follows: In Section \ref{sec:7Bedestruction}, we provide the derivation of the equations concerning the destruction of $^7$Be through interaction with a 2 MeV monochromatic line. We explore the possibility of generating such a line through a lukewarm 2-flavor superconducting (2SC-like) quark phase, where a fraction ($\eta_{\rm G}$) of the gluonic energy is converted into 2 MeV photons.
 The leakage phenomenon of the in-SQC QCD vacuum, involving a fraction ($f_{\rm V}$) of the SQC mass, is examined in Section \ref{sec:SQCs-DE}. Here, we present the resulting cosmology and discuss how it may offer a plausible resolution to the Hubble tension, a discrepancy in the measurement of the expansion rate of the universe.
In Section \ref{sec:SQCs-CDM}, we delve into the discussion of SQCs as a candidate for Cold Dark Matter (CDM). We analyze their properties and explore their viability as a constituent of the cosmic dark matter content.
The limitations and predictions of our model are outlined in Section \ref{sec:discussion}, offering insights into the boundaries and potential implications of our proposed framework.
Finally, we conclude the paper in Section \ref{sec:conclusion}, summarizing the key findings and discussing the broader implications of our model.

%
%
 
\section{Post-BBN $^7$Be destruction}
\label{sec:7Bedestruction}

In our model, the CSC phase produces gluon condensation (i.e. $M_{\rm LGB}c^2 \sim 4$ MeV mass LGBs) which decay to $E_0=M_{\rm LGB}c^2/2\sim$ 2 MeV mono-energetic photons.  
 To ensure that the energy of the 2 MeV line remains below the pair-creation threshold, we impose the condition 2 MeV < 12 MeV/$T_{\rm keV}$, which translates to $T_{\rm G} < 6$ keV. In other words, the QCs enter the color superconducting (CSC) phase and transform into SQCs during the radiation-dominated post-BBN era.
Henceforth, quantities labeled with the letter "G" pertain to values associated with the LGB-decay/photon-burst event. This event occurs when the age of the universe exceeds approximately 10 hours, corresponding to a redshift $z_{\rm G} = z(t_{\rm G}) < 2.5 \times 10^7$. 

The formation of LGBs takes place on a timescale comparable to that of hadronic processes when QCs enter the CSC phase. Subsequently, LGBs decay into photons almost instantaneously, with a timescale of $\tau_{\rm LGB}\sim 10^{-17}$ s, significantly shorter than the Hubble expansion timescale (as discussed in Section \ref{sec:the-2MEV-line}).

Subscripts "sG" and "eG" refer to the start and end, respectively, of this LGB-decay phase, denoted by $t_{\rm sG}$ and $t_{\rm eG}$. The duration of this phase is approximately equal to the LGB decay timescale, i.e., $t_{\rm eG}-t_{\rm sG} \sim \tau_{\rm LGB}$. Consequently, we have $z_{\rm sG} = z_{\rm eG} + \delta z$, where $\delta z$ is much smaller than $z_{\rm G}$. Effectively, this implies that $z(t_{\rm sG}) = z(t_{\rm G}) = z(t_{\rm eG})$, indicating that the redshifts at the start, during, and end of the LGB-decay phase are approximately equal.

 The photon production is a  delta function, $\delta(t-t_{\rm G})$, in our model at time $t=t_{\rm G}$. 
 As shown in Appendix \ref{appendix:PS15}, the  destruction rates for $^7$Be nuclei due to such a sudden release of mono-energetic $E_0$ 
photons  is
\begin{equation}
\label{eq:reduction}
\ln\left(\frac{Y_{\rm Be, eG}}{Y_{\rm Be, sG}}\right)\sim - \frac{n_{\gamma} (E_0, t_{\rm G})}{n_{\rm B}(t_{\rm G})}\times \frac{\sigma_{\rm Be}(E_0)}{\sigma_{\rm CS}(E_0)}\ ,
\end{equation}
where $Y$ is the abundance. Here,  $n_{\rm B}(t_{\rm G})$ is the universe's co-moving baryon 
 number density at $t_{\rm G}$ while $n_{\gamma} (E_0, t_{\rm G})$ is the co-moving number density of photons from the LGB
 decay.  The $^7$Be  photo-dissociation cross-section is $\sigma_{\rm Be}(E_0)$ and $\sigma_{\rm CS}(E_0)$
  the Compton scattering cross-section. A  reduction of $^7$Be by 2/3  imposes that the RHS in Eq. (\ref{eq:reduction}) be unity (i.e. $\sim -1.1$). 

 Let us define $\eta_{\rm DM, sG}$ as the initial total amount in mass of SQCs (the DM in our model) compared to baryons; i.e. before LGB decay.
The corresponding SQC co-moving number density is 
\begin{equation}
\label{eq:csqc-comoving-density}
n_{\rm sqc}^{\rm com.}(t_{\rm G})=\frac{\eta_{\rm DM, sG} n_{\rm B}(t_{\rm G})}{A_{\rm sqc}}\ ,
\end{equation}
 with  $A_{\rm sqc}$ the cluster's baryon number.  The emitted photons co-moving number density is then 
\begin{equation}
n_{\gamma} (E_0, t_{\rm G}) = n_{\rm sqc}^{\rm com.}(t_{\rm G})\times N_{E_0}=\frac{\eta_{\rm DM, sG} n_{\rm B}(t_{\rm G})}{A_{\rm sqc}}\times N_{E_0}\ ,
\end{equation}
with $N_{E_0}\sim A_{\rm sqc}\times ( \eta_{\rm G} m_{\rm p}c^2/E_0)$ the total number of $E_0$ photons emitted per SQC. 
Here, $\eta_{\rm G}$ is the fraction of the gluonic energy per equivalent proton rest-mass 
converted to the mono-chromatic line at energy $E_0$; $m_{\rm p}$ is the proton mass. 
 
 Eq. (\ref{eq:reduction}) becomes
\begin{align}
\label{eq:reduction2}
\ln\left(\frac{Y_{\rm Be, eG}}{Y_{\rm Be, sG}}\right)&\sim   - \eta_{\rm DM, sG} \frac{\eta_{\rm G} m_{\rm p}c^2}{E_0}\times \frac{\sigma_{\rm Be}(E_0)}{\sigma_{\rm CS}(E_0)}\\\nonumber 
&\sim -\frac{\eta_{\rm G}}{1-\eta_{\rm G}}\times \frac{\eta_{\rm DM, eG}m_{\rm p}c^2}{E_0}\times \frac{\sigma_{\rm Be}(E_0)}{\sigma_{\rm CS}(E_0)}\ ,
\end{align}
where, in order to get the last expression, we made use of the fact that the SQC total mass after conversion of gluonic energy
to photons is $\eta_{\rm DM, eG}= (1-\eta_{\rm G})\eta_{\rm DM, sG}$ or $\eta_{\rm DM, sG}=\eta_{\rm DM, eG}/(1-\eta_{\rm G})$.

  \begin{figure}  
\centering  
\includegraphics[scale=0.8]{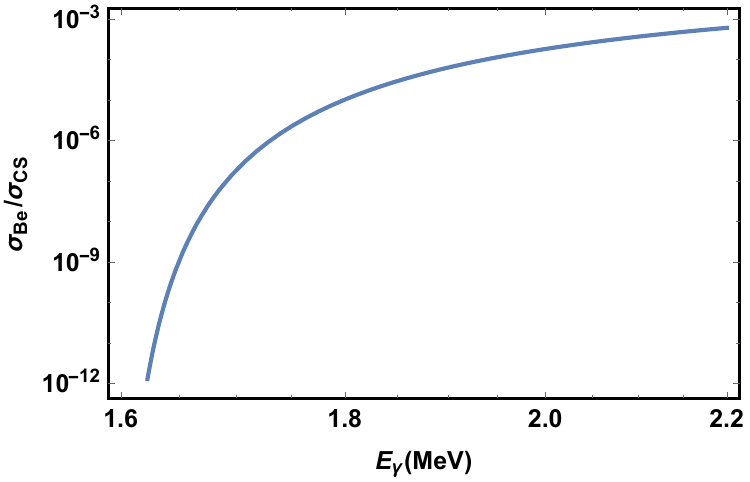}
\caption{The $\sigma_{\rm Be}/\sigma_{\rm CS}$ ratio versus photon energy. The $^7$Be photo-dissociation cross-section $\sigma_{\rm Be}$
is from equation III.8 in \cite{ishida_2014}. The Compton scattering cross-section, is given in Appendix (IV) in \cite{kawasaki_1995}.}
  \label{fig:ratio-sigmas} 
\end{figure}

The $^7$Be  photo-dissociation cross-section, $\sigma_{\rm Be}(E_0)$, is given by  Eq. (III.8) in \citet{ishida_2014}.
 The ratio of $\sigma_{\rm Be}(E_0)$ to Compton scattering cross-section, $\sigma_{\rm Be}(E_0)/\sigma_{\rm CS}(E_0)$, varies widely as shown in Figure \ref{fig:ratio-sigmas}. We see that for $2\ {\rm MeV} < E_0 < 2.2$ MeV we have 
 $3\times 10^{-4}< \sigma_{\rm Be}/\sigma_{\rm CS}< 8\times 10^{-4}$. If the SQC total mass after conversion of gluonic energy to photons is  the observed CDM amount, $\eta_{\rm DM, eG}\sim 5$, then 
 $5m_{\rm p}c^2/E_0\sim 2.4\times 10^3$. In this case, the $^7$Li problem is solved if
  $0.65< \frac{\eta_{\rm G}}{1-\eta_{\rm G}}< 1.56$ meaning when 

\begin{equation}
\label{eq:etaG}
0.4 < \eta_{\rm G}< 0.6\ ,
\end{equation}
which requires that on average $\sim$ 50\% (i.e. $\eta_{\rm G}\sim 0.5$) of the SQCs gluonic energy is converted to $\sim 2$ MeV photons.

 As explained later in Section \ref{sec:discussion}, the value of $\eta_{\rm G}$ can be lower (less gluonic energy shed by SQCs) while still solving the $^7$Li puzzle. This is because in addition to losing gluonic energy via LGB decay, SQCs will lose more during the leakage of its QCD vacuum so that
 if $\eta_{\rm DM, 0}\sim 5$ in today's universe\footnote{The subscript ``0" refers to values at redshift $z=0$.} then $\eta_{\rm DM, eG}$ must have been higher due to leakage (see first bullet point in Section \ref{sec:discussion}).
 
 It is crucial to emphasize the requirement for a narrow band emission, preferably a mono-energetic photon source, in order to maximize the number of photo-dissociated $^7$Be nuclei (as described by Eq. (\ref{eq:reduction})). Without such a narrow emission, there would not be sufficient gluonic energy available from the SQCs - which serve as the Dark Matter (DM) in our model - to achieve the desired 2/3 reduction. In this context, the decay channel of LGBs presents an appealing mechanism for generating such a narrow emission (see Appendix \ref{appendix:LGBs}). The feasibility of this mechanism can be tested in ongoing experiments (as indicated in bullet point \#\ref{itemCBM}  in Section \ref{sec:discussion}).

\subsection{The mono-chromatic $\sim 2$ MeV line}
\label{sec:the-2MEV-line}

\begin{figure}
\centering  
\includegraphics[scale=0.5]{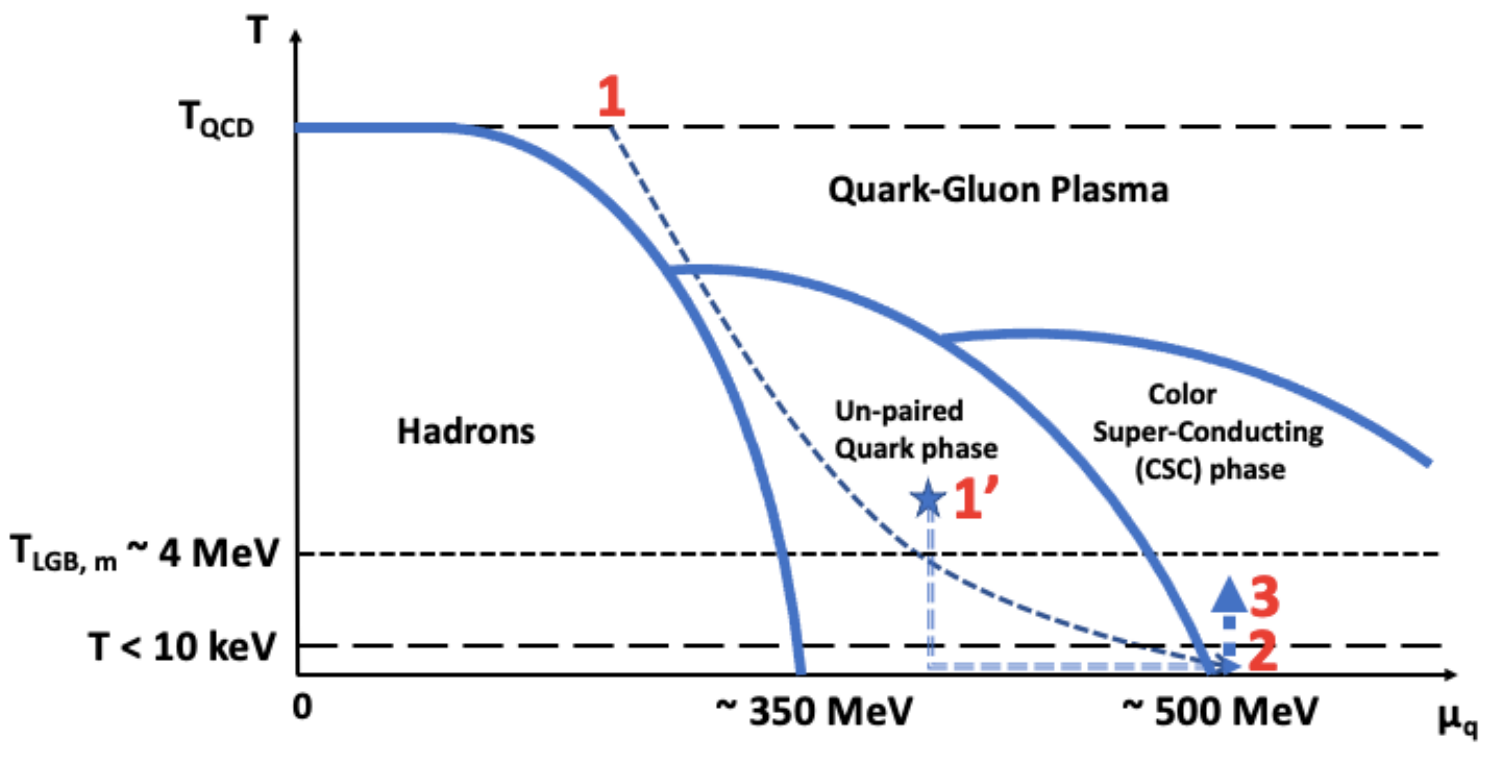}
\caption{A possible QCD phase diagram; temperature versus quark chemical potential. The dashed curve (``1 to 2")  
depicts the cooling path of the quark cluster (formed at ``1" with $T_{\rm QCD}\sim 150$ MeV) traversing the unpaired phase
and entering the CSC phase, at keV temperature, from the low density phase (``2"). A first-order line separates the unpaired phase
from the CSC phase releasing heat (the vertical ``2 to 3" arrow). A SQC gets heated to a temperature not exceeding the LGB melting temperature
$k_{\rm B}T_{\rm sqc, m}\sim M_{\rm LGB}c^2$ (see Section \ref{sec:the-2MEV-line}). The $\sim 4$ MeV LGBs decay to the 2 MeV narrow photon line via ${\rm LGB}\rightarrow \gamma+\gamma$. ``$1^\prime$" shows the initial state of a NS core making its way to ``2" via cooling and compression (see Section \ref{sec:discussion} and Appendix \ref{appendix:LGBs}).
}
\label{fig:transition-to-CSC}
\end{figure}

 The QCD phase diagram is complex  (e.g. \cite{ruster_2004,alford_2008,baym_2018}) and in principle  one
  cannot exclude  the existence of a CSC phase where gluon condensation would yield electromagnetically unstable $X$ ``particles" (here LGBs), as required in our model. Figure \ref{fig:transition-to-CSC} shows a hypothetical phase diagram with the dashed line depicting a possible trajectory leading a quark cluster from its state at birth (in an unpaired phase)  to the CSC phase.  We require an unpaired phase at chemical potential $\mu<\mu_{\rm csc}$ which bridges the  hadronic phase and the CSC phase.  I.e. the CSC phase is accessed,  at low-$T$, from low density unpaired phase with a first-order line separating the two phases.
Once in the CSC phase, a cluster becomes a SQC and produces LGBs in the radiation-dominated post-BBN epoch at keV temperatures. 

  In this paper, we will use the neutral 2SC phase as a reference CSC phase. It has the interesting property of
converting a percentage ($\eta_{\rm G}=3/8$) of its gluonic energy to LGBs (i.e. gluonic condensation)  at low temperature
 with a subsequent decay to a mono-chromatic line via the ${\rm LGB}\rightarrow \gamma+\gamma$ channel (see Appendix 
 \ref{appendix:LGBs} for details).  In the 2SC phase, an 
LGB with mass $M_{\rm LGB}c^2\sim 4$ MeV  would decay to two $E_0=(M_{\rm LGB}c^2/2)\sim 2$ MeV  photons on timescales of $\tau_{\rm LGB}\sim 10^{-17}$ s.  We set the CSC quark chemical potential at $\mu_{\rm csc}= 500$ MeV with
a corresponding  number density $n_{\rm csc}= \mu_{\rm csc}^3/\pi^2\sim  10^{39}$ cm$^{-3}$.
 Thus the density inside\footnote{Not to be confused with the SQC co-moving density given in Eq. (\ref{eq:csqc-comoving-density}).} an SQC is $n_{\rm sqc}=n_{\rm csc}$ which is about ten times nuclear saturation density when it enters the CSC phase at  time $t_{\rm G}$. 

Once the cluster crosses into the CSC phase, the first order transition proceeds on hadronic timescales. 
 During the transition,  and because of latent heat released, a SQC gets heated to a temperature $k_{\rm B}T_{\rm sqc} \sim \Delta_{\rm csc}^2/\mu_{\rm csc}$ where $\Delta_{\rm csc}$ is the CSC superconducting gap; here, $\Delta_{\rm csc}$ and $\mu_{\rm csc}$
 are in units of MeV (see Eq. \ref{eq:Tsqc}). 
  LGBs cannot form at temperatures exceeding the melting temperature which is of the order of the LGB's rest-mass energy; $k_{\rm B}T_{\rm LGB, m}\sim M_{\rm LGB}c^2$.  Thus  we must ensure that $T_{\rm sqc} < T_{\rm LGB, m}$  in addition to $E_0=E_{\rm LGB}/2< 2.2$ MeV where  
 the 2.2 MeV upper limit keeps the line below the  Deuteron photo-ionization threshold (see Section \ref{sec:introduction}). 
 In general for $\Delta_{\rm csc}^2/\mu_{\rm csc} < 2$ MeV, or $\Delta_{\rm csc} < 31.6\ {\rm MeV}\times (\mu_{\rm csc}/500\ {\rm MeV})^{1/2}$, these conditions are satisfied. $E_{\rm LGB}\simeq M_{\rm LGB}c^2\sim 4$ MeV is achieved  for a reasonable range in $\mu$ as shown in Figure \ref{fig:LGB-mass}. The width of the $\sim 2$ MeV line is given in Eq. (\ref{eq:width}) in Appendix \ref{appendix:LGBs}, and for typical values, it is expected to be $< 0.1$ MeV.

\subsection{The SQC size and baryon number}
\label{sec:SQC-size}

 To first order, the SQC's photon mean-free-path is $\lambda_{{\rm csc}, \gamma}=1/(n_{\rm csc}\sigma_{{\rm q}\gamma})\sim 10^2$ fm$/n_{\rm csc, 39}$ with $\sigma_{{\rm q}\gamma}\sim 10^{-28}\ {\rm cm}^2$ the Thomson cross-section for photon scattering off up and down quarks.  
By setting a typical SQC radius to be $R_{\rm sqc, thin}\sim \lambda_{{\rm csc}, \gamma}$, the SQC is optically thin to all photons including  the 2 MeV photons. Being transparent to electromagnetic radiation of all energies, the SQC makes a good DM candidate (that is still within the standard model of particle physics) 
 with a corresponding baryon number  $A_{\rm sqc}\sim 10^6$.

\begin{figure}
\centering
\includegraphics[scale=0.6]{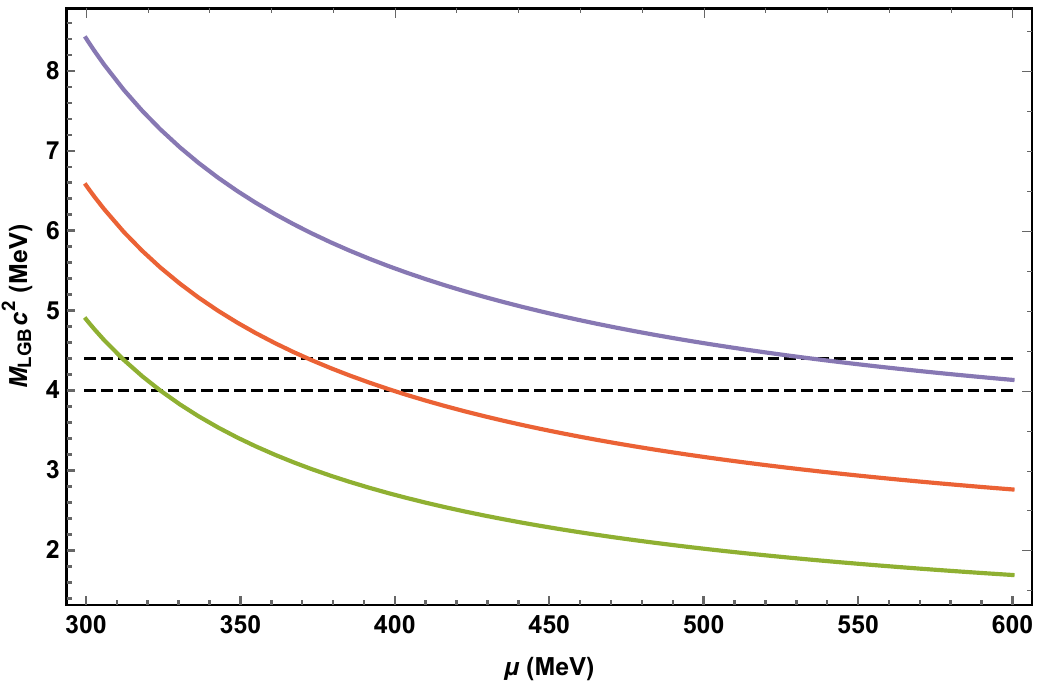}
\caption{LGB mass versus quark chemical potential ($\mu$)  for $\Lambda_{\rm QCD}= 245$ MeV (the scalar parameter of QCD; see Section \ref{sec:the-2MEV-line} and Appendix \ref{appendix:LGBs} for details). The curves from top to bottom are for $\Delta/\mu =0.1,0.09,0.08$, respectively.  The two dashed lines show the $4.0 \le M_{\rm LGB}c^2 ({\rm MeV}) \le 4.4$ range yielding a mono-chromatic photon line in the $2.0 \le E_0 ({\rm MeV})\le 2.2$ range.  Using $\Lambda_{\rm QCD}\sim 340$ MeV expected using the usual renormalization scheme with 3 quark flavours, we get $M_{\rm LGB}\sim 4$ MeV for $\Delta \sim 0.05\mu$ (see Eq. (\ref{appendix:lambdac})).}
  \label{fig:LGB-mass}
\end{figure}

%
%

\section{SQCs and Dark Energy (DE)}
\label{sec:SQCs-DE}

Both hadrons and SQCs exist as states within QCD, but they exhibit different symmetries associated with their respective vacuum states. Hadrons, as discussed in \cite{brodsky_2012}, are characterized by confinement in their vacuum state, while SQCs are influenced by the formation and decay of LGBs  to photons, leading to distinct symmetries. In Appendix \ref{appendix:confined-SUC(2)}, we discuss the confinement of the $SU_c(2)$ phase within SQCs, where the vacuum state is speculated to undergo a tunnelling process, referred to as ``leakage," into the external trivial vacuum. This tunnelling phenomenon is triggered by instabilities resulting from various mechanisms, including electromagnetic gluonic decay, such as the formation and subsequent decay of light gluonic bound states into photons in the CSC phase.

Although we chose to focus on the 2SC phase due to its simplicity, it is important to note that any CSC quark phase capable of converting gluonic energy into approximately 2 MeV photons and initiating a transition to the trivial vacuum could be equally effective in our proposed framework.
This innovative concept offers an intriguing explanation for DE and establishes a cosmological framework that has the potential to address the Hubble tension while still yielding a universe consistent with the $\Lambda$CDM model at low redshifts.

 The total amount of confined $SU_c(2)$ QCD vacuum energy stored in the SQCs 
  is $\rho_{\rm QCD}^{\rm vac.}V_{\rm sqc}^{\rm tot.}$ with $\rho_{\rm QCD}^{\rm vac.}$  the density of the QCD condensates
   and $V_{\rm sqc}^{\rm tot.}$ the total volume occupied by the SQCs which is constant in time (and assumed to not change on LGB decay). 
  "Leakage" means that the density inside the SQC decreases from
 \begin{equation}
 \label{eq:rhocscsqneG}
 \rho_{\rm sqc, eG}=(1-\eta_{\rm G})\rho_{\rm csc}\ ,
 \end{equation}
  at $t=t_{\rm G}$ (at the end of the LGB phase denoted by ``eG") to 
 \begin{align}
 \label{eq:rhocscsqn0}
  \rho_{\rm sqc, 0}&=(1-\eta_{\rm G})\rho_{\rm csc}-\rho_{\rm QCD}^{\rm vac.}\\\nonumber
   & = \rho_{\rm csc}\times (1-\eta_{\rm G}-f_{\rm V})\\\nonumber
  &= \rho_{\rm sqc, eG}\times \left( 1- \frac{f_{\rm V}}{1-\eta_{\rm G}}\right)\ ,
  \end{align}
   at redshift $z = 0$ with the assumption that the leakage 
  timescale is a fraction of the age of the universe $t_0\sim 13.5$ Gyrs (see below).  The parameter $f_{\rm V}= \rho_{\rm QCD}^{\rm vac.}/\rho_{\rm csc}$
   is a measure of the contribution of the QCD condensates to the SQC rest-mass energy; naturally $f_{\rm V}< 1-\eta_{\rm G}$. 
   At the start of the LGB phase (denoted by ``sG"), we have $ \rho_{\rm sqc, sG}=\rho_{\rm csc}$.

  We can get a rough estimate of the time $t_{\rm tun.}$ (the e-folding time) it would take the in-SQC QCD vacuum to 
leak into the exterior space-time trivial QCD vacuum. A  rigorous calculation would follow for example \citet{coleman_1977} and is beyond the scope of this paper. We consider instead a simple tunnelling problem across a square barrier with height given by the expectation value of the QCD vacuum and width $L\sim R_{\rm sqc}$. 
 The corresponding tunnelling probability is then
 $P_{\rm tun.}\sim e^{-R_{\rm sqc}/\delta}$ where $\delta$ is the penetration depth (e.g. \cite{tannoudji_2006}).
  The tunnelling timescale is $t_{\rm tun.}\simeq (R_{\rm sqc}/c)\times (1/P_{\rm tun.})$
  giving us 
  \begin{equation}
  t_{\rm tun.}\sim  \frac{\delta}{c}\times 
 x e^{x}\ ,
  \end{equation}
  where $x=R_{\rm sqc}/\delta$; the tunnelling timescale is highly sensitive to $x$. With $\delta$ of the order of a Fermi (which is not unrealistic), solutions with  tunnelling timescales on the order of a billion years ($t_{\rm tun.}\sim$ Gyr) require $R_{\rm sqc}$ to be approximately 88 fm. If $R_{\rm sqc}$ exceeds 90 fm, the tunnelling timescale $t_{\rm tun.}$ becomes longer than the age of the universe, $t_0$. Furthermore, the size of the SQC cannot be too small for leakage to occur on astrophysical timescales, ensuring that $t_{\rm G} << t_{\rm tun.}$. Therefore, an optimal range for the SQC radius is found to be $85 < R_{\rm sqc} ({\rm fm}) < 88$, which is consistent with our model. It is worth noting that in our model, optically thin SQCs with $R_{\rm sqc} < 10^2/n_{\rm csc, 39}$ fm are required to address the cosmological $^7$Li problem. Furthermore, the size of the SQC cannot be too small for leakage to occur on astrophysical timescales, ensuring that $t_{\rm G} << t_{\rm tun.}$. Therefore, an optimal range for the SQC radius is found to be $85 < R_{\rm sqc} ({\rm fm}) < 88$, which is consistent with our model. It is worth noting that in our model, optically thin SQCs with $R_{\rm sqc} < 10^2/n_{\rm csc, 39}$ fm are required to address the cosmological $^7$Li problem.

 The time evolution of the density  inside the SQC we can then write as
 \begin{align}
 \label{eq:rhosqc-evolution}
  \rho_{\rm sqc}(t)&=(1-\eta_{\rm G})\rho_{\rm csc}-\rho_{\rm QCD}^{\rm vac.}(1-e^{-(t-t_{\rm G})/t_{\rm tun.}})\\\nonumber
   &= \rho_{\rm csc} \left(1- \eta_{\rm G}-f_{\rm V}(1-e^{-(t-t_{\rm G})/t_{\rm tun.}})\right)\\\nonumber
   &= \rho_{\rm sqc, eG} \left(1- \frac{f_{\rm V}}{1-\eta_{\rm G}}(1-e^{-(t-t_{\rm G})/t_{\rm tun.}})\right)\ .
 \end{align}
 The equation above incorporates the key parameters in our cosmology namely, $\rho_{\rm csc}, \eta_{\rm G}, f_{\rm V}$ and $t_{\rm tun.}$
 which are all fundamentally related to QCD.
 
  The DM density in our model is the SQC density $\rho_{\rm sqc}(t)$ averaged over the volume of the universe. 
  Its   time evolution, with $V_{\rm univ.}(t)$ being the Hubble volume at time $t$,  is 
 
 \begin{equation}
\label{eq:rhoDM}
\rho_{\rm DM}(t)  =
\begin{cases}
  \frac{\rho_{\rm sqc}(t) V_{\rm sqc}^{\rm tot.}}{V_{\rm univ.}(t)}\ & {\rm if} \quad t > t_{\rm G}\ ({\rm or}\ z < z_{\rm G})\\
   \frac{\rho_{\rm csc} V_{\rm sqc}^{\rm tot.}}{V_{\rm univ.}(t)}\ & {\rm if} \quad t \le t_{\rm G}\ ({\rm or}\ z \ge z_{\rm G})\ ,
 \end{cases}   
\end{equation}
with the resulting cosmology analyzed in Appendix \ref{appendix:our-cosmology}.  
 
\begin{figure}
  \centering
  \includegraphics[scale=0.4]{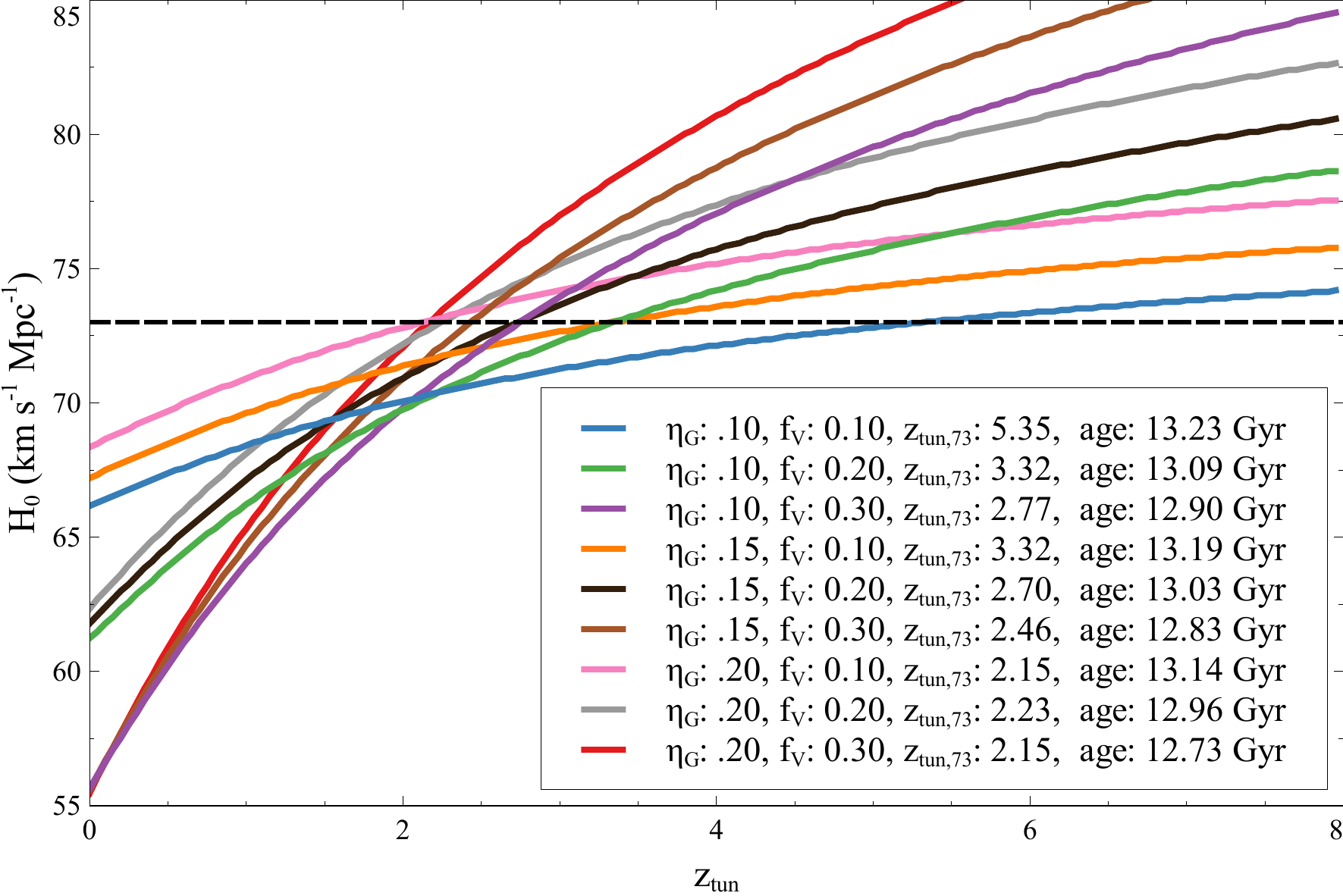}
 \caption{The Hubble constant $H_0$ as a function of $z_{\rm tun.}$ (the leakage characteristic redshift) in our model 
 for different values of $\eta_{\rm G}$ and $f_{\rm V}$. The $H_0=73$ km s$^{-1}$ Mpc$^{-1}$ value  is shown as the horizontal line. 
  The resulting age of the universe  is shown for each case and is younger than the $\Lambda$CDM universe.}
  \label{fig:H0}
  \end{figure}

  Our model offers a resolution of the Hubble tension (see \cite{kamionkowski_2022} for a recent review and references therein)
  and can be understood as a consequence of a CDM universe converting into a $\Lambda$CDM universe at $z_{\rm tun.}$. 
   Figure \ref{fig:H0} shows that $H_0\sim73$ km s$^{-1}$ Mpc$^{-1}$ can be obtained for a range in $\eta_{\rm G}$ and $f_{\rm V}$ values with a 
  leakage characteristic redshift  $2 < z_{\rm tun.} < 6$ (i.e. $1 < t_{\rm tun.} ({\rm Gyr})< 3.3$).   It is both noteworthy and encouraging that the timescale $t_{\rm tun.}$, which is tightly constrained by the size of the SQC at approximately 100 Fermi, coincides with the timescale required to address the Hubble tension. This convergence underscores the interconnection of these two phenomena.
  
  Our cosmology yields a universe which is younger than the $\Lambda$CDM universe with an age of $\sim 13$ Gyrs with $\eta_{\rm G}$ and $f_{\rm V}$
   of the order of tens of percents each.    It remains to be shown whether our cosmology is in agreement with other cosmological data and measurements which the flat $\Lambda$CDM model explains extremely well (see e.g. metrics and tests suggested in \cite{schoneberg_2022}).
 Furthermore, we caution that the details of how the QCD vacuum mixes with the space-time vacuum and how it evolves while
    preserving flatness remains to be understood.  Nevertheless, being a vacuum the   ``DE" component in our model 
  should obey an equation-of-state with parameter $w=-1$.

%
%

\section{SQCs as cold dark matter (CDM)}
 \label{sec:SQCs-CDM}
 
 Evidence for CDM is abundant but its nature  remains unknown despite many theoretical 
 investigations  and dedicated experiments which have yet to detect any associated particle (e.g.
  \cite{garrett_2011,schumann_2015,bertone_2018,kisslinger_2019,oks_2021,arbey_2021}).
  In our model, the $^7$Li problem is solvable if the total mass in SQCs is of the order of the measured
  CDM value. In addition, SQCs are colorless, cold, optically thin, electrically neutral and would make
  an ideal CDM candidate if they decouple from the strong force (or interact  minimally with baryons)  following leakage.

  We have put forward the argument that SQCs that have undergone tunnelling would decouple from hadrons, resulting in their interaction being solely gravitational in nature. Consequently, these tunnelled SQCs would elude detection in current Dark Matter (DM) experiments. In the pre-LGB era, QCs should possess a sufficiently extended lifespan, allowing them to persist into the post-BBN era at $z_{\rm G}$ without experiencing any growth in size due to interactions with hadrons. The rate at which non-tunnelled SQCs interact with hadrons depends on the fraction of SQCs that have survived until the present time. For instance, for a tunnelling timescale of $t_{\rm tun.}\sim 1$ Gyr, this survival fraction is $e^{-t_0/t_{\rm tun.}}\sim 10^{-6}$, and it drops significantly to $e^{-t_0/t_{\rm tun.}}\sim 10^{-12}$ for $t_{\rm tun.}\sim 0.5$ Gyr. Below, we employ the term ``baryon" specifically to refer to non-QC baryons (e.g. protons), allowing us to examine the interactions of QCs and SQCs with hadrons during these two distinct epochs.
   
    \subsection{The pre-LGB era} 
    
    In the pre-LGB (i.e. pre-SQC at $z > z_{\rm G}\sim 2.5\times 10^7$) epoch, QCs should interact with baryons.  The QC-baryon reaction rate is  $\sigma_{\rm qc-B}n_{\rm B} v_{\rm qc-B}$ with $n_{\rm B}(T)=\eta_{\rm B}n_{\gamma}(T)$ 
the baryon's number density and $n_{\gamma}(T)\sim 6.2\times 10^{32}$ cm$^{-3}\times T_{\rm MeV}^3$;  $\eta_{\rm B}\simeq 6.1\times 10^{-10}$ is the baryon-to-photon ratio. The QC-baryon interaction cross-section
is $\sigma_{\rm qc-B}$ while the QC-baryon relative velocity $v_{\rm qc-B}= \sqrt{k_{\rm B}T/m_{\rm p}}\simeq 10^{-3/2}c\times T_{\rm MeV}^{1/2}$ is dominated by that of baryons; $c$ is the speed of light and $m_{\rm p}$ the proton mass.
 For the QC not to grow in size in the pre-LGB era (i.e. up to $z_{\rm G}$) we impose 
$\int_{\mu s}^{hours} \sigma_{\rm qc-B}n_{\rm B} v_{\rm qc-B} dt < A_{\rm qc}$. With $t = 1\ {\rm s}\times  T_{\rm MeV}^{-2}$ we have $d (t/1\ {\rm s})=-2 T_{\rm MeV}^{-3}dT_{\rm MeV}$ so that $7.2\times 10^{32} \sigma_{\rm qc-B}\times \int_{T_{\rm BBN, keV}}^{\rm T_{\rm QCD, MeV}} T_{\rm MeV}^{1/2} dT_{\rm MeV}< A_{\rm qc}$ gives 

\begin{equation}
\label{eq:sigma-qc-B}
\sigma_{\rm qc-B} < 10^{-36}\ {\rm cm}^2\times A_{\rm qc}\ .
\end{equation}
I.e. $\sigma_{\rm qc-B}/M_{\rm qc}< 10^{-12}\ {\rm cm}^2/{\rm g}$ with $M_{\rm qc}=A_{\rm qc}m_{\rm p}$.

The above ensures that the QC will neither grow in size in the pre-LGB era nor interact with baryons in the subsequent periods where baryons are more diffuse. 
We can express the effective cross-section in terms of the geometric one as $\sigma_{\rm qc-B}  =\pi R_{\rm qc}^2\times (\tau_{\rm crossing}/\tau_{\rm conv.})$. Here,  $\tau_{\rm crossing}=R_{\rm qc}/v_{\rm sqc-B}$   is the baryon QC crossing time and  $\tau_{\rm conv.}$  the conversion timescale of a baryon to the CSC phase
(i.e. for a baryon to be absorbed by the QC).   The ratio $\tau_{\rm crossing}/\tau_{\rm B-qc, conv.}$ accounts for the fraction of time the baryon spends inside the QC. With $A_{\rm qc}=\frac{4\pi}{3}R_{\rm qc}^3 n_{\rm qc}$, the condition in Eq. (\ref{eq:sigma-qc-B}) translates to $\tau_{\rm B-qc, conv.}> 10^{-13}\ {\rm s}$; here $n_{\rm qc}\sim 10^{39}$ cm$^{-3}$ is the QC's density.  

The fact that the conversion timescale $\tau_{\rm conv., B-qc}$ exceeds the timescales associated with strong interactions provides support for the notion that the QC state is distinct from the hadronic phase. Alternatively, it suggests that the conversion process may involve two steps, where the baryon first undergoes deconfinement, possibly requiring an injection of energy, before transitioning into the unpaired quark phase characteristic of QCs. Another interpretation is that quarks within a hadron, where chiral symmetry is broken, perceive the QC as a distinct and separate phase of quark matter even prior it becomes an SQC. This intriguing aspect highlights the complexity and richness of the underlying physics involved in the conversion process and the distinct properties of QC and SQC phases.

 \subsection{The post-LGB era}
 
  From this point forward, when we refer to SQCs, we are referring to the clusters that did not undergo tunnelling. The SQCs that have undergone tunnelling, on the other hand, constitute the Dark Matter (DM) component within our proposed model. To impose a constraint on the SQC-baryon interaction cross-section $\sigma_{\rm SQC-B}$, we will utilize the measured lower limit on the proton lifetime ($\tau_{\rm p} > 3.6\times 10^{33}$ years; \cite{superK_2002}). This lower limit on the proton lifetime provides valuable information about the strength of the SQC-baryon interaction and allows us to further refine our model.
   
The SQC reaction rate per baryon is $\sigma_{\rm sqc-B}  n_{\rm sqc} v_{\rm sqc-B}$ and $v_{\rm sqc-B}$
the SQC-baryon relative velocity. The SQC co-moving number density (not to be confused with the SQC density $n_{\rm csc}\sim 10^{39}$ cm$^{-3}$) is 
 
 \begin{equation}
\label{eq:sigma2}
n_{\rm sqc}\sim \frac{\rho_{\rm DM}}{A_{\rm sqc}m_{\rm p}}\times e^{-t_0/t_{\rm tun.}}\ ,
\end{equation}
where $\rho_{\rm DM}\sim 0.36$ GeV cm$^{-3}$ is the local DM density \cite{sofue_2020} which is that of the tunnelled SQCs in our model.
Imposing $\sigma_{\rm sqc-B}  n_{\rm sqc} v_{\rm sqc-B}< 1/\tau_{\rm p}$ and using $v_{\rm sqc-B}\sim 240$ km s$^{-1}$ we arrive at  

\begin{equation}
\label{eq:sigma-sqc-B}
\sigma_{\rm sqc-B} <  10^{-48}\ {\rm cm}^2\times A_{\rm sqc} \times e^{t_0/t_{\rm tun.}}\ .
\end{equation}
I.e. $\sigma_{\rm sqc-B}/M_{\rm sqc}< (10^{-24}\ {\rm cm}^2/{\rm g})\times e^{-t_0/t_{\rm tun.}}$ with $M_{\rm sqc}=A_{\rm sqc}m_{\rm p}$.

Consistent with expectations, the aforementioned analysis reveals that the efficiency of tunnelling, indicated by the presence of fewer remaining SQCs today, directly correlates with the permissible range for the SQC-baryon interaction cross-section $\sigma_{\rm sqc-B}$. Specifically, a more efficient tunnelling process allows for a higher allowable value of $\sigma_{\rm sqc-B}$. This relationship underscores the significance of the tunnelling mechanism in influencing the interaction dynamics between SQCs and baryons within our model.
In terms of the baryon-to-SQC conversion timescale $\tau_{\rm B-sqc, conv.}$ (to be differentiated from the conversion of a baryon to an QC
above), and with $\sigma_{\rm sqc-B} = \pi R_{\rm sqc}^2\times (\tau_{\rm crossing}/\tau_{\rm B-sqc,conv.})$ we get
$\tau_{\rm B-sqc,conv.} > \frac{12.5\ {\rm s}}{n_{\rm csc, 39}}\times e^{-t_0/t_{\rm tun.}}$. 

  In section \ref{sec:SQCs-DE}, we have established that a cosmological model capable of resolving the Hubble tension and simultaneously yielding the observed age of the universe necessitates a tunnelling timescale of approximately one gigayear (Gyr), corresponding to an exponential suppression factor of $e^{-t_0/t_{\rm tun.}}\sim 10^{-6}$. Consequently, this implies that $\sigma_{\rm sqc-B} < 10^{-42}\ {\rm cm}^2\times A_{\rm sqc}$, or equivalently, that the conversion timescale $\tau_{\rm B-sqc,conv.}$ exceeds $10^{-5}$ seconds. 
  
 Intuitively, assuming $\sigma_{\rm sqc-B}< \sigma_{\rm qc-B}$ would imply $e^{-t_0/t_{\rm tun.}}> 10^{-12}$ or $t_{\rm tun.}> 0.4$ Gyr
 which is also consistent wth our cosmology. We caution though that making direct comparisons between the QC and SQC states may not be justified. 
  We should add that if SQC vacuum leakage is not a simple tunnelling process then it can deviate from a pure exponential process (e.g. \cite{kanno_2012}) which could affect
  our findings here. This is left as an avenue for future research. 
    
Furthermore, we can employ neutron stars (NSs) to impose even more stringent constraints on the SQC-baryon cross-section by studying the interactions between SQCs and baryonic matter within a NS.
  If $\sigma_{\rm sqc-B} n_{\rm NS} R_{\rm NS}> A_{\rm SQC}$, the SQC slows down inside the neutron star and is captured. 
 For typical neutron star parameters, such as $n_{\rm NS}\sim 10^{38}$ cm$^{-3}$ and $R_{\rm NS}=10^6$ cm, the condition for no capture implies that $\sigma_{\rm sqc-B} < 10^{-44}\ {\rm cm}^2\times A_{\rm sqc}$. Referring to Eq. (\ref{eq:sigma-sqc-B}), this condition is always satisfied if $e^{-t_0/t_{\rm tun.}}>10^{-4}$, indicating that neutron stars will not be converted through SQC capture.  For a NS formed at time $t_{\rm NS}$, the mass of SQCs it accumulates over the Hubble time can be estimated as $\left( \int_{t_{\rm NS}}^{t_0} \pi R_{\rm NS}^2 \frac{\rho_{\rm DM}}{A_{\rm sqc}m_{\rm p}}e^{-t/t_{\rm tun.}} v_{\rm sqc-NS} dt\right) \times A_{\rm sqc}m_{\rm p}\sim 10^{11}\ {\rm g}\times t_{\rm tun., Gyr}\times (e^{-t_{\rm NS}/t_{\rm tun.}}- e^{-t_0/t_{\rm tun.}})$. Here, we assume a SQC-neutron star relative velocity of $v_{\rm sqc-NS} \sim 10^7$ cm s$^{-1}$ and a representative local DM density of $\rho_{\rm DM}\sim 0.1$ GeV cm$^{-3}$. Thus, even if a certain amount of SQC is captured, it is not evident whether it would be sufficient to convert a neutron star.

Due to the likelihood that the leakage timescale exceeds the typical timescale for the formation and subsequent collapse of structures during the early universe, we anticipate that primordial halos will be predominantly composed of non-leaked SQCs that formed after the Light Glueball (LGB) era. Baryonic matter will gravitationally accrete onto these halos, but as demonstrated earlier, the diffuse baryons will not interact with either leaked or non-leaked SQCs. However, we anticipate a critical change or "phase transition" in the growth of structures after the time $t=t_{\rm tun.}$ This is because: (i) a significant fraction of the SQCs' mass is converted into Dark Energy (DE), and (ii) the SQCs would have decoupled from hadrons. Understanding the precise impact of these effects on subsequent structure growth requires more advanced investigations beyond the scope of this paper. Only state-of-the-art Cold Dark Matter (CDM) simulations, which incorporate the leakage of SQC mass as detailed in Appendix \ref{appendix:our-cosmology}, can provide insights into this aspect of our model.
 
 To summarize, SQCs present a viable candidate for Cold Dark Matter (CDM) and may play a role in the formation of both mini-halos and larger structures in the universe, as indicated in previous studies \cite{navarro_1996, abel_2002}. Once halos form and reach a state of virialization, SQCs are expected to interact most strongly with stars, particularly with matter at high densities such as neutron stars (NSs). However, if the tunnelling timescale $t_{\rm tun.}$ is a small fraction of the age of the universe, as suggested earlier, then SQCs would have decoupled from hadrons by the time compact stars begin to form (as discussed in Section \ref{sec:SQCs-DE}). In other words, even the densest stars, such as neutron stars, are unlikely to undergo conversion through capture of SQCs. However, if the cores of neutron stars can access the Color Superconducting Quark matter (CSC) phase through phenomena like cooling and mass accretion, it could have intriguing implications for astrophysics (as shown in equation (\ref{eq:compactness})). This possibility opens up new avenues for research in understanding the behavior and properties of neutron stars within the framework of our model.

%
%

 \section{Discussion} 
 \label{sec:discussion}

 Here we briefly discuss some limitations and distinctive features of our model and leave these as future investigations.
 
 \begin{enumerate}
 
 \item {\bf The evolution of the SQC and of the DM densities}:   We first recall, from Eq. (\ref{eq:rhosqc-evolution}), that the density inside an SQC evolves from $\rho_{\rm csc}$ to $(1-\eta_{\rm G})\rho_{\rm csc}$
after the LGB phase. leakage further decreases the SQC density to an asymptotic value of $(1-\eta_{\rm G}-f_{\rm V}) \rho_{\rm csc}$  at $t>> t_{\rm tun.}$ (which we associate with $z=0$). The minimum possible value of an SQC  would be given by the 
bare quarks masses or, $\sim 90$\% of their original mass. 
 However, we find $\eta_{\rm G}$ and $\eta_{\rm V}$ to be of the order of tens of percents each (see Section \ref{sec:SQCs-DE}), and an SQC would lose a fraction of its mass to DE over the age of the universe. 
Meaning that the DM content in today's universe is a fraction of that in the pre-BBN era according to our model.
  
 Averaging the total SQC mass over the Hubble volume gives us the DM density in our model. 
 The ratio of total amount of DM to that of the baryonic matter  evolves from a maximum value of $\eta_{\rm DM, sG}$
 before LGB decay to  $\eta_{\rm DM, eG}=(1-\eta_{\rm G})\eta_{\rm DM, sG}$
 after the LGB decay to photons  just before leakage starts.  At full leakage, which we associate with $z=0$,
  the ratio is 
  \begin{equation}
  \label{eq:etas}
  \eta_{\rm DM, 0}=(1-\eta_{\rm G}-f_{\rm V})\eta_{\rm DM, sG}= (1-\frac{f_{\rm V}}{1-\eta_{\rm G}})\eta_{\rm DM, eG}\ .
  \end{equation} 
  Assuming $\eta_{\rm DM, 0}\sim 5$ as measured today and extrapolating back to the LGB era, it means that 
  the amount of DM content is larger than what we used in Section \ref{sec:7Bedestruction} where we set $\eta_{\rm DM, eG}=\eta_{\rm DM, 0}\sim 5$
  (i.e. when taking $f_{\rm V}=0$).
 
\item {\bf The $^7$Li problem revisited}:    The solution to the cosmological $^7$Li problem presented in Section \ref{sec:7Bedestruction}
did not take into account further loss of gluonic energy due to in-SQC vacuum leakage following LGB decay as discussed above. 
From Eq. (\ref{eq:etas}), $\eta_{\rm DM, sG}= 5/(1-\eta_{\rm G}-f_{\rm V})$ and when 
 plugged in Eq. (\ref{eq:reduction2}),  we get 
 
 \begin{equation}
\label{eq:etaG2}
0.65 < \frac{\eta_{\rm G}}{(1-\eta_{\rm G}-f_{\rm V})}< 1.56\ .
\end{equation}
 
 Eq. (\ref{eq:etaG}) becomes
 
 \begin{equation}
\label{eq:etaG3}
0.4\times (1-f_{\rm V}) < \eta_{\rm G} < 0.6\times (1-f_{\rm V})\ .
\end{equation}
I.e. a smaller percentage of the gluonic energy of the SQCs converted to LGBs could resolve the $^7$Li problem.  
For $f_{\rm V}=0$ we recover the $\eta_{\rm G}$ values we arrived at in Section \ref{sec:7Bedestruction} while  
 for $f_{\rm V}> 2/5$ we get $\eta_{\rm G}< 3/8$ which is  less than the maximum $\eta_{\rm G}\sim 3/8$  expected from the 2SC-phase. 

  The ratio between the DE density and the DM energy density at $z=0$ is (from Eq. (\ref{eq:rhocscsqn0})), 
  \begin{equation}
  \frac{\rho_{\rm DE}}{\rho_{\rm DM}}=\frac{\rho_{\rm QCD}^{\rm vac.}}{\rho_{\rm sqc, 0}}= 
  \frac{\rho_{\rm QCD}^{\rm vac.}}{(1-\eta_{\rm G})\rho_{\rm csc}-\rho_{\rm QCD}^{\rm vac.}}=\frac{f_{\rm V}}{1-\eta_{\rm G}-f_{\rm V}}\ .
  \end{equation}
  To have $\rho_{\rm DE}\sim 2 \rho_{\rm DM}$ as measured in today's universe requires $f_{\rm V}\sim \frac{2}{3}(1-\eta_{\rm G})$ in which case 
  Eq. (\ref{eq:etas}) gives $0.18 < \eta_{\rm G} < 0.34$ and $0.44 < f_{\rm V} < 0.55$. 
    On the other hand, the resolution of the Hubble tension suggests that a favored parameter combination lies around $(\eta_{\rm G},f_{\rm V})\sim (0.1,0.1)$, as illustrated in Figure \ref{fig:H0}. This discrepancy could potentially be mitigated by relaxing some of the approximations made in our cosmological model or by adopting a more realistic approach to the tunnelling process, moving beyond the simple assumption of an exponential decay.

\item \label{itemNSs}{\bf The CSC phase and Neutron stars}:  Figure \ref{fig:transition-to-CSC} shows a suggested pathway, 
      starting at point ``$1^\prime$", a NS core could take to enter the CSC phase
     since conversion following SQC capture is suppressed (see Section \ref{sec:SQCs-CDM}).  A NS born with (or which acquires through evolution)
a core in the unpaired phase could transition to the CSC phase by a sequence of cooling  (to keV temperature via
the URCA process; e.g. \cite{paczynski_1972}) and compression (to $\mu_{\rm csc}=500$ MeV via mass accretion). 
 Take a NS with a core making up a fraction $\eta_{\rm c}$ of the total mass. The energy released from conversion of gluonic condensation (e.g. LGBs) to photons is 
    $(\eta_{\rm c} M_{\rm NS}/m_{\rm p})\times (\eta_{\rm G} m_{\rm p}c^2) = \eta_{\rm c}\eta_{\rm G}M_{\rm NS}c^2$. Comparing this  to
    the NS binding energy $\frac{3}{5}GM_{\rm NS}^2/R_{\rm NS}$, we conclude that NSs with compactness parameter 
     \begin{equation}
     \label{eq:compactness}
    \frac{M_{\rm NS, \odot}}{R_{\rm NS, 6}} < 0.11 \times \frac{\eta_{\rm c}}{0.1}\times \frac{\eta_{\rm G}}{0.1}\ ,
    \end{equation} 
    may be completely obliterated in the process; the NS mass and radius are in units of solar mass, $M_{\odot}$, and $10^6$ cm, respectively.
    NSs  with higher compactness parameter would loose mass leaving behind a pure CSC core. 
    In this latter case, the conversion  to a CSC star puts a constraint on $\rho_{\rm csc}$ due to  the black hole limit $2GM_{\rm NS}/c^2 < R_{\rm csc}$ with $R_{\rm csc}$ the radius of the CSC star. With   $\rho_{\rm csc}R_{\rm csc}^3\sim\rho_{\rm NS} R_{\rm NS}^3$ this gives  $\rho_{\rm csc}< 10^{16}\ {\rm g\ cm}^{-3}/M_{\rm NS, \odot}^{2}$
   which is  consistent with the $\mu_{\rm csc}\sim 500$ MeV (i.e. a 2SC-like phase) adopted in our model. 
        Thus, if some NSs follow a path as suggested in  Figure \ref{fig:transition-to-CSC}, the resulting photon fireball
          may have interesting implications to explosive astrophysics. 
 
  \item   \label{itemDecoupling}{\bf SQC-hadron decoupling}:   Following leakage and loss of more gluonic energy,  it is not unreasonable to assume that the quarks within the SQC become undressed and should in principle  decouple (or at least experience some level of decoupling) from the strong interaction. They would still rely on gluons to remain bound while exhibiting minimal interaction with hadrons.    We speculate that DM and hadrons may represent separate phases of quark matter within the framework of QCD, characterized by distinct vacuum properties.
   If our model is correct, it allows a unique connection between cosmology and the properties of the QCD vacuum in the CSC quark phase.
 An estimate of the parameter $f_{\rm V}$ from cosmological observations  may be an indication of the contribution
 of the QCD condensates in CSC to the SQC mass which may have implications, albeit an indirect one, to the mass of hadrons. The exact details of the SQC-Hadron decoupling remain to be worked out.

\item \label{itemCBM} {\bf SQCs and LGBs in today's detectors}:  SQCs (the DM in our model; see Section \ref{sec:SQCs-CDM}) would interact only gravitationally and would 
 thus evade detection in current DM  experiments.   Instead we propose that our model can be tested by experiments which can access the 2SC phase at temperatures below the LGB melting temperature; i.e. at
 $< M_{\rm LGB} c^2\sim 4$ MeV (see Appendix \ref{appendix:LGBs}). The Compressed Baryonic Matter experiment (CBM) at FAIR
 explores the QCD phase diagram in the region of high baryon densities
 (representative of neutron star densities) and moderate temperatures using high-energy nucleus-nucleus collisions  (e.g. \cite{ablyazimov_2017,senger_2020}).  We note that the 2SC phase carries $M_{\rm LGB}c^2\sim 4$ MeV LGBs  (i.e. can solve the cosmological $^7$Li problem) at baryonic density as low as a few times that of nuclear matter (see Figure \ref{fig:LGB-mass}) and that LGBs form on strong interaction timescales and decay  to MeV photons on timescales of $\sim 10^{-17}$ s. Thus, experiments such as the CBM at FAIR could potentially be used to detect the photons  from the LGB decay.  
 
 The width of the line can be found in Eq. (\ref{eq:width}) in Appendix \ref{appendix:LGBs}, and for typical values, it is expected to be $< 0.1$ MeV. If this width were to be measured, it would provide a direct estimation of the 2SC dielectric constant, which could then be used to infer the $\Delta/\mu$ ratio and consequently determine $\Delta$ (as per Eq. \ref{eq:width}). However, it remains unclear whether and how the CBM experiment could explore the temperature regime relevant to our study. More significantly, the 2 MeV line may be overshadowed by the anticipated high low-energy electromagnetic background in high-energy ion-ion collisions. Even if the background were to diminish as the QGP transitions into the 2SC phase and subsequently cools down to the LGB phase ($T<4$ MeV), detecting the 2 MeV line would likely prove challenging. Calculating the strength of the LGB 2 MeV line is beyond the scope of this paper.

\item {\bf The stability of the CSC phase}: Our findings seem to hint at the standard neutral 2SC phase (adopted in our theoretical framework) 
   as the unspecified CSC phase. However, the 2SC phase may be unstable at small superconducting gap ($\Delta$)  values due to the
  mismatch in the up and down quarks chemical potential \cite{huang_2004}. It suggests that either the 2SC phase is stable 
  in the regime of chemical potential ($\mu$) and $\Delta$ values we used or that another stable 2SC-like phase exits
  in nature and remains to be identified. 
    A 2SCus phase, which has u-s pairs instead of u-d pairs, is also a candidate.  This phase, however, faces a similar challenge in that the u and s Fermi momenta are split apart by the strange-quark mass (combined with electrical neutrality; e.g. \cite{alford_2005}).  The Color-Flavor-Locked phase may be one candidate if the strange-quark effective mass is small or alternatively a crystalline quark phase
    if strange-quark mass is heavier (e.g. \cite{cao_2015}). However, these phases do not possess properties that allow for the conversion of gluonic energy to photons.  Despite its limitations, the u-d 2SC phase remains as one of the candidate phases for dense quark matter (adopted in our model), pending the determination of its exact phase structure.
    
 Another possibility are  collective excitations (i.e. phonons) of SQCs which would  decay directly to photons via phonon-to-photon conversion channels.  Mono-chromatic photons of energy $1.59\ {\rm MeV}<  E_{\gamma}< 22.2$ MeV would result if the SQC radius were $ 544.6\ {\rm fm} < R_{\rm sqc}\sim h c/E_{\gamma}\sim 753.6$ fm. Physically, the decaying particle is not an LGB but  is instead  a long-wavelength mode which is resonant with the quark cluster containing it.  In this scenario, higher resonant modes should also be excited and these could affect the very sensitive D abundance which is not desired. Superconducting strings \cite{haber_2018} is an interesting avenue to explore in this context. If such domain walls could form during the cosmic QCD phase, one could imagine a scenario where they
would evolve into 2SC-dominated strings in the post-BBN era. It remains to be shown that LGBs as described in our model could form in this case.

   \item \label{itemFormation}{\bf Matter-Antimatter annihilation and SQC size}: We hypothesize that each quark cluster is born with an anti-matter deficit of $\eta_{\rm B}\simeq 6.1\times 10^{-10}$
  meaning that there is one extra baryon per  $\eta_{\rm B}^{-1}$ quark-antiquark pairs; $\eta_{\rm B}$ is the baryon-to-photon ratio.  After annihilation
(on timescales of $1/n_{\rm csc} \sigma_{\rm annih.} \sim 10^{-13}$ s with $\sigma_{\rm annih.}\sim$ mbarn), 
 a cluster has only baryons left in it; this assumes that annihilation does not destroy the cluster and instead it
 reduces it into a pure baryon cluster of radius $R_{\rm sqc, f}\sim R_{\rm sqc, thin}$ where ``f" stands for final.
  Here,  $R_{\rm sqc, thin}\sim 10^2$ fm is the typical size of a SQC set by the photon mean-free-path in the
  2SC phase. In other words, we claim that the maximum size of the ``shrapnel" of the annihilated much bigger parent cluster to be 
 of the order of $R_{\rm sqc, thin}$ (see Section \ref{sec:SQC-size}). In this case, a cluster's birth radius can be obtained from $n_{\rm csc} R_{\rm sqc, thin}^3/n_{\rm csc} R_{\rm sqc, b}^3= \eta_{\rm B}$
 which gives $R_{\rm sqc, b}\sim 10^5$ fm$/n_{\rm csc, 39}$.    Some constraints and implications to consider in the future include: 
   (i) Annihilation should also yield pions. These would decay on weak-interaction timescales and if their mean-free-path  turns out to be much smaller than that of photons they may affect the $\sim 10^2$ fm SQCs; (ii) While cluster formation
   (when the universe has aged such that its temperature is in the tens of MeV)  is followed rapidly by annihilation, we must avoid re-creation of matter-antimatter pairs. I.e. ensure that  pair-creation timescales exceed the Hubble expansion timescale; (iii) In the framework 
    we outline here, SQCs would require a formation mechanism which is different from that of the much larger $A>>A_{\rm qc}$
     cosmic strange-quark nuggets \cite{witten_1984} (which require a first-order QCD phase transition) and Axion quark nuggets \cite{zhitnitsky_2003} 
     (which require their co-existence with Axions).  Additionally, the mechanism by which these nuggets can convert their stored gluonic energy into $\sim 2$ MeV photons is a serious limitation.
     
 \end{enumerate}

%
%
 
 \section{Conclusion}
 \label{sec:conclusion}
 
 We have proposed that a color superconducting (CSC) phase of lukewarm QCD matter  could offer a non-exotic  solution to the cosmological $^7$Li, the CDM  and  the Dark Energy enigmas.  The narrow 2 MeV photon line which destroys $^7$Be in the radiation-dominated post-BBN epoch we attribute to gluonic condensation  (i.e. light glueballs or LGBs) in the CSC phase and its electro-magnetic decay modes (Section \ref{sec:7Bedestruction}).  The detailed properties of the CSC phase  remain to be scrutinized although a neutral 2SC-like phase  with a superconducting gap $\Delta < 0.1\mu$  is hinted at.  
 
 CDM, according to our model, consists of  colorless, charge neutral, optically thin cosmic quark clusters in the CSC phase (SQCs) with $R_{\rm sqc}\sim 100$ fm in size  and  baryon number $A_{\rm sqc}\sim 10^6$. They  decouple from hadrons and interact only gravitationally thus evading detection in current DM experiments.   If SQCs could be produced in experiments such as the Compressed Baryon Experiment at FAIR (see bullet point \#\ref{itemCBM} in Section \ref{sec:discussion}), they could be detected via the MeV photons from $LGB\rightarrow\gamma+\gamma$ decay giving support to our model. 
 
 The decoupling of SQCs from hadrons is due to leakage of the in-SQCs vacuum into the trivial 
 vacuum of the exterior space-time which yields DE.  As leakage proceeds, our cosmology gradually transitions from a non-DE to a DE ($\Lambda$CDM-like) universe at moderate redshift while allowing for a possible resolution of the Hubble tension (Section \ref{sec:SQCs-DE}). 
  It is crucial to recognize the significance of having optically thin SQCs with a size on the order of 100 Fermi in resolving the $^7$Li problem. This requirement directly sets the timescale for the tunnelling process to be on the order of a Giga-year. The connection between the Fermi scales, which govern particle interactions, and the astrophysical scales suggests a  fundamental connection between the $^7$Li problem and Dark Energy (DE)
 unique to our model.

 Our model does not introduce new physics to solve the $^7$Li, DM and DE problems but instead makes use of still uncertain properties of QCD phases and its vacuum properties.   We are not the first to discuss a connection between QCD vacuum and cosmology. It has been argued based on empirical properties of hadrons that confinement is a pre-requisite for retaining condensates inside hadrons, which then largely eliminates the problem of the smallness of the cosmological constant.     On the other hand, our model,  introduces the concept of tunnelling of the in-SQC vacuum into the exterior trivial vacuum, and the decoupling of SQCs from hadrons (see Section \ref{sec:SQCs-DE}). This distinctive proposition raises the intriguing possibility that DM and hadrons could represent separate phases of quark matter within the framework of QCD, characterized by distinct vacuum properties
 that may turn out to have other useful physical applications.
  

\funding{R.O., D.L. and N.K.  acknowledge the support of the Natural Sciences and Engineering Research Council of Canada (NSERC).
     P.J. is supported by a grant from the Natural Science Foundation PHY-2310003. }

\dataavailability{No new data were generated or analyzed in support of this research.}

\acknowledgments{We thank R. Rapp for interesting discussions on various aspects of the paper. 
     R.O. thanks P. Serpico for brief comments on Appendix A. We thank the referees for their insightful comments and suggestions which improved the quality and rigour of our paper.}
     
\conflictsofinterest{The authors declare no conflicts of interest.}
     

%
%

\appendixtitles{no} 
\appendixstart
\appendix

 \section{Pedagogical framework}
 \label{appendix:pedagogical}

 \subsection{A Quantum Chromodynamics (QCD) phase diagram}

\begin{figure}
  \centering
  \includegraphics[scale=0.4]{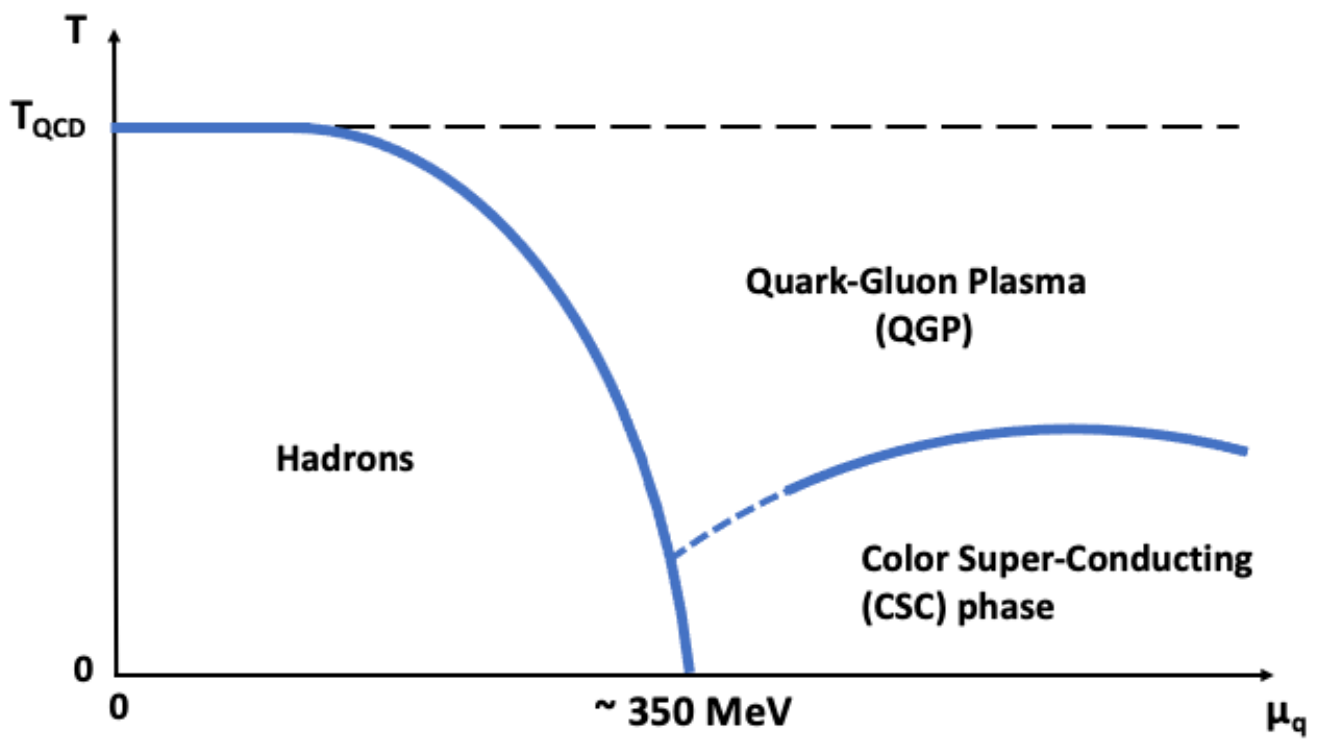}
 \caption{A simplified QCD phase diagram in the density-temperature plane; the density is represented by the quark chemical potential $\mu_{\rm q}$. The hadronic phase exists at temperature $T<T_{\rm QCD}\sim 150$ MeV
 and chemical potential $\mu_{\rm q}< \mu_{\rm h}\sim 350$ MeV. The dashed line in the diagram represents unexplored phases that could potentially exist within the QGP phase before reaching the CSC phase.}
  \label{fig:QCD-Phase-Diagram}
  \end{figure}

 QCD is the theory governing the strong nuclear force. Figure \ref{fig:QCD-Phase-Diagram} shows a simplified 
  QCD phase diagram in the T-$\mu_{\rm q}$ (temperature vs the quark chemical potential) plane featuring three distinct phases. The 
  chemical potential is related to the number density by  $n= \mu^3/\pi^2$. 
   At low temperatures and densities is the confined phase, where quarks and gluons are bound within color-neutral hadrons, such as protons and neutrons. In this phase, quarks are unable to exist freely as individual particles. As the temperature and/or density increases, the QCD phase diagram exhibits a transition to a deconfined phase called the quark-gluon plasma (QGP). In the QGP phase, quarks and gluons move more freely and are not confined within hadrons.  Lattice simulations have shown that there is 
    a smooth transition (a cross-over) between the confined hadronic phase and the deconfined quark-gluon plasma phase (see
 \cite{guenther_2021} for a recent review). Unlike a first-order phase transition, where there is a distinct jump in thermodynamic quantities at a critical point, a cross-over is characterized by a gradual change in the system's behavior. This means that there is no precise temperature at which the transition occurs, but rather a temperature range over which the transition takes place.

At higher densities (or chemical potential) and lower temperatures is the color superconductivity (CSC) phase. This phase is characterized by the condensation of Cooper pairs of quarks, similar to how electrons form pairs in traditional superconductors. A CSC is characterize by its superconducting gap ($\Delta_{\rm csc}$  in the tens of MeV range)  which is the energy required to break a Cooper pair. The vacuum ground state energies in the CSC and hadronic phases are drastically different. 
The formation of Cooper pairs alters the QGP, leading to changes in the gluon spectrum and dispersion relations. It has been demonstrated that the CSC  is the ground state of quark matter at asymptotically large densities (e.g. \cite{alford_2008}). 

The simplest form of CSC is known as two-flavor color superconductivity (2SC), which arises in QCD when quarks of two different flavors (usually up and down quarks) form Cooper pairs due to the attractive interaction mediated by the strong force.  The Cooper pairs in the 2SC phase are made of the red and green quarks only, and the blue quarks do not participate in pairing. Only five out of a total eight gluons become massive while the other three gluons remain massless and do not interact with the paired red and green quarks nor with the gapless blue quarks. 
These 3 gluons obey their own gluon-dynamics and form their own condensate made of bound states of pure gluons similar to how mesons are bound states of quarks and antiquarks.  This gluon condensate is unstable electromagnetically  and emits photons.  
The decay process of the electrically neutral condensate into photons occurs through its interaction with virtual quark loops, which carry the electric charge (see Appendix \ref{appendix:LGBs}).
This electromagnetic decay plays a crucial role in resolving the cosmological Lithium-7 problem within our model, as explained in the following description.

   \subsection{Our model in a nutshell}
  
  The cosmological lithium-7 ($^7$Li) problem relates to a significant discrepancy of a factor of 3 between the predicted abundance of cosmological beryllium-7 ($^7$Be) and the lower measured abundance of $^7$Li in the universe. In this context, we present a proposal for a non-exotic electromagnetic solution within the standard model of particle physics to address the cosmological $^7$Li problem, a crucial element in the theory of Big Bang nucleosynthesis (BBN). Our model utilizes properties of deconfined quark matter to offer a potential solution to the enigma of $^7$Li. Additionally, we explore the interconnectedness of this solution with the problems of dark energy (DE) and dark matter (DM), all of which highlight our limited understanding of the quark-gluon plasma (QGP) and the phases of quark matter as described in QCD.
  
 At the core of our model lies the conversion of the gluonic energy (i.e. of the gluons) of superconducting deconfined quark matter into a narrow 2 MeV line, capable of selectively destroying primordial $^7$Be without impacting other products or the physics of the cosmic microwave background (CMB). One key assumption in our model posits that, in addition to the known hadrons (protons, neutrons, and mesons) formed during the QCD phase transition in the early universe, assemblies of deconfined quark matter, referred to as quark clusters (QCs), can also form. These QGP clusters have a size of approximately 100 Fermi (and are thus optically thin to photons) and are composed of up, down quarks, and gluons in proportions that makes them charge neutral. Inside a QC, the density reaches approximately ten times the nuclear saturation density, which is on the order of $10^{39}$ cm$^{-3}$ (equivalently, $\mu_{\rm q} \sim 500$ MeV). This high-density environment corresponds to a baryon number, with $A_{\rm qc} \sim 10^6$.

The second key assumption in our model is that these QCs transition into a color superconducting state (CSC) in the post-BBN era, when the universe is a few hours old, at a redshift of $z < \sim 2.5\times 10^7$.  The 2-flavor color superconductivity (2SC) phase specifically involves two flavors of quarks, typically up and down quarks as seen in QCs. When QCs reach his phase and become superconducting (referred to as SQCs), a percentage of their gluons form Light Glueballs (LGBs). Glueballs are bound states of gluons, analogous to how mesons are bound states of quarks and antiquarks. LGBs are considered to be relatively light, with a mass in the MeV range, compared to other possible glueball states with masses in the GeV range. LGBs are electromagnetically unstable, and by appropriately selecting 2SC parameters, they can decay into a monochromatic $\sim 2$ MeV line, capable of selectively destroying primordial $^7$Be while leaving other BBN products and CMB physics unaffected. 

To resolve the $^7$Li problem, approximately 20-30\% of the gluonic energy in SQCs needs to be converted into a monochromatic 2 MeV line. Interestingly, the total mass of SQCs required for this solution corresponds to the observed DM in the universe. SQCs, being cold, colorless, charge-neutral, and optically thin to all photon energies, present an attractive DM candidate that is deeply intertwined with the enigma of $^7$Li. 
We would like to emphasize  the necessity of a narrow band emission (preferably a mono-energetic photon source) to optimize the number of photo-dissociated $^7$Be nuclei (see Eq. (\ref{eq:reduction})). Otherwise, there will not be enough gluonic energy from the SQCs (the DM in our model) for the 2/3 reduction. The LGB decay channel is an attractive mechanism for such a line (see Appendix \ref{appendix:LGBs}) and may be tested in current heavy-ion collision experiments (see bullet point \#\ref{itemCBM}  in Section \ref{sec:discussion})

The nature of the QCD ground state continues to be a topic of active research and debate. There are many possible forms of the ground state wavefunction (in a many-body variational approach). Even when comparing the many ground state using free energies one can never be certain that the true ground state is found (see discussion in  \cite{alford_2001b}).  In principle, the vacuum state within SQCs after the formation and decay of LGBs can potentially exhibit different symmetries compared to the vacuum state of confined quark matter in hadrons; see Appendix \ref{appendix:confined-SUC(2)}. This brings us to the third assumption in our model 
which has implications for DE. By assuming that the in-SQC QCD vacuum becomes unstable following LGB decay, we allow it to "leak" into the trivial vacuum of the surrounding space-time via quantum tunnelling (Appendix \ref{appendix:confined-SUC(2)}). This "leakage" process behaves as a cosmological constant and occurs late in the evolution of the universe (at a redshift $z_{\rm tun.}$ of a few units), leading to a transition to a $\Lambda$CDM-dominated era (see Figure \ref{fig:H0}). This "leakage" effectively separates the pre-$\Lambda$CDM and $\Lambda$CDM phases and offers a potential resolution to the Hubble tension (see Section \ref{sec:SQCs-DE}).

Through our efforts to resolve the $^7$Li problem, we have uncovered a solution that effectively tackles the challenges of DE and DM, thus revealing the intrinsic connection of these critical cosmological issues within our proposed model.  For example, the in-SQC vacuum would be stable to decay if its gluonic sector were stable (as in hadrons). The cosmological $^7$Be nuclei act as absorbers of LGB decay products (with the 2 MeV photons ideal in this scenario) and avoid affecting BBN and CMB properties. Without addressing the $^7$Li, DM, and DE problems simultaneously, our model would not hold together. Despite its assumptions and limitations, it is important to reiterate that our framework operates within the standard model of particle physics but in an unexplored regime of QCD, emphasizing the need for further investigation and empirical validation.

\section{Post-BBN $^7$Be destruction}
\label{appendix:PS15}

 In \citet{poulin_2015} (and references therein), the photon-generating particle  $X$ decays exponentially on timescale $\tau_{\rm X}$ and forms at early time in the universe at $t$  $(<<\tau_{\rm X})$.
  In our model, the $X$ particle (i.e. the LGB) forms and  decays to a narrow $E_0\sim$ 2 MeV photons nearly instantaneously in the
 radiation-dominated post-BBN era at time $t_{\rm G}$ when the quark cluster enters the CSC phase; here ``G" stands for gluons
 as a reference to the LGB era (see Section \ref{sec:7Bedestruction}). Thus,   we have a delta function
production at time $t_{\rm G}$; $\delta(t-t_{\rm G})$. The source injection term (e.g. Eq. (2) in \citet{poulin_2015} and Section IV in \citet{kawasaki_1995}) becomes

\begin{equation}
\label{eq:appendix:source}
S(E_{\gamma},t)= n_{\gamma} (E_{\gamma}, t)\times \delta(t-t_{\rm G})\times \delta(E_{\gamma}-E_0)\ ,
\end{equation}
with $n_{\gamma} (E_{\gamma}, t)$ the co-moving photon density at time $t$; $\delta(E_{\gamma}-E_0)$ is the spectral shape factor for a monochromatic line at $E_0$.
 
Making use of Eqs. (4) and (5) in \citet{poulin_2015} with the destruction of element $A$ only (i.e. $\gamma+A\rightarrow P$) gives
the abundance as 

\begin{equation}
\label{eq:appendix:integral}
\frac{dY_A(t)}{dt} = -Y_A(t) \int dE_{\gamma}\sigma_{\gamma+A\rightarrow P}(E_{\gamma})\times \frac{S(E_{\gamma},t)}{\Gamma(E_{\gamma},t)}\ ,
\end{equation}
where $\Gamma(E_{\gamma},t)$ and $\sigma_{\gamma+A\rightarrow P}(E_{\gamma})$ are the photon  interaction rate and  the photo-dissociation cross-section, respectively.  The interaction rate is for all processes including Bethe-Heitler pair-creation on nuclei, double photon pair-creation, scattering off  thermal background photons and Compton scattering.  Red-shifting is negligible because the rates of electromagnetic interactions are faster than the cosmic expansion rate. 

Inserting Eq. (\ref{eq:appendix:source}) into Eq. (\ref{eq:appendix:integral})  and  integrating over energy gives

\begin{equation}
\frac{d\ln{Y_A(t)}}{dt} =- n_{\gamma} (E_0, t)\times \delta(t-t_{\rm G}) \frac{\sigma_{\gamma+A\rightarrow P}(E_0)}{\Gamma(E_0,t)}\ .
\end{equation}

This assumes a single scattering ($N_{\rm scat}=1$, as in our model) is enough to make the
$E_0$ photon no longer dissociate a $^7$Be nucleus. Otherwise $\Gamma(E_0, t)$ should be divided by $N_{\rm scat}$.
 
 Dissociation starts at redshift $z_{\rm sG}$ and ends abruptly at 
 redshift $z_{\rm sG}=z_{\rm eG}+\delta z$ with $\delta z<<z_{\rm sG}$; subscripts ``eG" and ``sG" stand for end and start, respectively, of the photon burst from LGB decay in the post-BBN era.  
After integration over a very small time $t$ with $t_{\rm sG}<t_{\rm G}$ and $t_{\rm eG}>t_{\rm G}$ but $z(t_{\rm sG})=z(t_{\rm eG})=z(t_{\rm G})$ one obtains:

\begin{equation}
\label{appendix:eq:reduction}
\log{\left(\frac{Y_{\rm Be, eG}}{Y_{\rm Be, sG}}\right)} =- n_{\gamma} (E_0, t_{\rm G})\ \frac{\sigma_{\rm Be}(E_0)}{\Gamma(E_0,t_{\rm G})}\ ,
\end{equation}
where  $n_{\gamma} (E_0, t_{\rm G})$ is the co-moving number density
of $E_0$ photons from LGB decay. Here,  $\sigma_{\rm Be}(E_0)\equiv \sigma_{\gamma+A\rightarrow P}(E_0)$ while 
$Y_{\rm Be, sG}\equiv Y_A(t_{\rm sG})$ and $Y_{\rm Be, eG}\equiv Y_A(t_{\rm eG})$ 
 are the corresponding  $^7$Be abundances.

With $E_0<2.2$ MeV, Bethe-Heitler pair-creation on nuclei  (see Eq.(4-356) in \cite{lang_1999})
 is negligible and so is pair creation off the CMB background photons when $T_{\rm CMB}$ is a few keV or less; the interaction rate
  is dominated by Compton scattering (CS). Thus 
  $\Gamma(E_0,t_{\rm G})\simeq \Gamma_{\rm CS}(E_0,t_{\rm G})= n_{\rm e}(t_{\rm G}) \sigma_{\rm CS}(E_0)$ with the CS cross-section 
   given in Appendix (IV) in \citet{kawasaki_1995}.
  Furthermore, because at $T_{\rm CMB}< 20$ keV positrons are negligible (e.g. Appendix C in \cite{svensson_1990}), 
     the total electron density is  of the order of the baryon number density, $n_{\rm e}(t_{\rm G})\sim n_{\rm B}(t_{\rm G})$. 
The  destruction rates for $^7$Be nuclei due to a sudden release of mono-energetic $E_0$ photons at $t_{\rm G}$ becomes
\begin{equation}
\label{appendix:eq:reduction2}
\ln\left(\frac{Y_{\rm Be, eG}}{Y_{\rm Be, sG}}\right)\sim - \frac{n_{\gamma} (E_0, t_{\rm G})}{n_{\rm B}(t_{\rm G})}\times \frac{\sigma_{\rm Be}(E_0)}{\sigma_{\rm CS}(E_0)}\ .
\end{equation}

\section{Light glueballs (LGBs) in the 2SC phase}
\label{appendix:LGBs}

The spectrum in the 2SC state is made of 5 massive gluons with a mass
of the order of the gap $\Delta$, 3 massless gluons  and gapless up and down quarks in the direction 3 (blue) of color; in this
appendix we use QCD natural units with $\hbar=c=1$.
The 3 massless gluons in the 2SC phase do not interact with the gapless blue quasi-particles and the quasiparticles from the green and red paired
quarks decouple from the low-energy confined $SU_c(2)$ phase. The 3 massless gluons in the 2SC phase bind, or are confined,  into LGBs 
 when the temperature is below the melting temperature $T_{\rm LGB, m}\sim M_{\rm LGB}$ (\cite{ouyed_2001} and references therein).
I.e. LGBs  melt when the temperature exceeds the confinement value \cite{sannino_2002}.  

The LGB mass  at $T\le T_{\rm LGB, m}$ is $M_{\rm LGB}\sim \hat{\Lambda}_{\rm c}$ with $\bar{\Lambda}_{\rm c}$ 
the intrinsic scale associated with the
$SU_c(2)$ theory. It is  the confining scale of the $SU_c(2)$ gluon-dynamics in 2SC and  when $\Delta << \mu$, the one loop relation gives \cite{rischke_2001}  
\begin{equation}
\label{appendix:lambdac}
\bar{\Lambda}_{\rm c}=\Delta \exp{\left(-\frac{2 \sqrt{2}\pi}{11}\times \frac{\mu}{g_{\rm s}\Delta}\right)}\ ,
\end{equation}
 with $g_{\rm s}=\sqrt{4\pi \alpha_{\rm s}}$ and $\alpha_{\rm s}= \frac{12\pi}{(33-2n_{\rm f}) \ln{(\mu^2/\Lambda_{\rm QCD}^2)}}$ the $SU_c(3)$  coupling constant evaluated at $\mu$; $\Lambda_{\rm QCD}$ is the scalar parameter of QCD
and $n_{\rm f}$ the number of relevant quark flavors. For $\Delta \sim 0.1\mu$  and for $\Lambda_{\rm QCD}= 245$ MeV (expected from a pure gluonic theory), we get
$M_{\rm LGB}\sim 4$ MeV for a range in quark chemical potential as shown in Figure \ref{fig:LGB-mass}.   Using $\Lambda_{\rm QCD}\sim 340$ MeV expected using the usual renormalization scheme with 3 quark flavours, we get $M_{\rm LGB}\sim 4$ MeV when $\Delta \sim 0.05\mu$.

Once created, LGBs are stable against strong interactions but not with respect to electromagnetic processes.
The two-photon decay mechanism of the electrically neutral LGB was estimated in \cite{ouyed_2001} based on the saturation of the electromagnetic trace anomaly at the effective Lagrangian level. In other words, the coupling between the LGBs, which dominate the energy-momentum tensor at low energies, and two photons occurs through virtual quark loops which carry the electric charge. The decay occurs on timescale $\tau_{\rm LGB}\sim 5.5\times 10^{-14}\ {\rm s}\times (M_{\rm LGB}/\rm MeV)^{-5}$ \cite{ouyed_2001}.

The pairing energy density released during the unpaired-to-2SC phase transition is $Q_{\rm pairing}=\mu^2\Delta^2/\pi^2$ (e.g. \cite{alford_2001,shovkovy_2004}).
 Thus during the transition, to a first approximation a SQC gets heated to a temperature 
 \begin{equation}
 \label{eq:Tsqc}
 T_{\rm sqc}= \frac{Q_{\rm pairing}}{n_{\rm csc}}\sim \frac{\Delta^2}{\mu}\ .
 \end{equation}
LGBs would not melt during the unpaired-to-CSC heating phase if $\Delta^2/\mu < M_{\rm LGB}$ allowing for the
$LGB\rightarrow \gamma+\gamma$ to occur (see Section \ref{sec:the-2MEV-line}).

 The LGBs within an SQC move with the same velocity, denoted as $v$, as the underlying gluons. 
 The broadening of the $E_0=M_{\rm LGB}/2$ line can then be expressed as
 
 \begin{equation}
 \label{eq:width}
 \frac{\Delta E_0}{E_0} =\frac{\frac{1}{2}M_{\rm LGB}v^2}{E_0}= v^2= \frac{1}{\epsilon}\ . 
  \end{equation}
 The last equality in the equation above, means that  the velocity is determined by the dielectric constant $\epsilon$ of the 2SC medium given as
$\epsilon=1+\frac{g_{\rm s}^2}{18\pi^2}\times \left(\frac{\mu}{\Delta}\right)^2$ \cite{rischke_2001}.  We see that $\Delta<<\mu$ yields $\epsilon>>1$ (or $v<<1$) which ensures a narrow line.  The $v<<1$ regime is also the reason that the LGBs are not dynamically important in the 2SC phase because their mass scale as $v^{-3/2}$ relative to the in-vacuum case.

When $\Delta < 0.1\mu$, $\epsilon > 10$ and $\Delta E_0/E_0< 0.1$. For instance for $\Lambda_{\rm QCD}\sim 340$ MeV, where $\Delta \sim 0.05\mu$ is required to get $M_{\rm LGB}\sim 4$ MeV, we have $\epsilon\sim 30$ or $\Delta E_0\sim 0.067$ MeV. It also implies that the range of $E_0$ falls between 2.0 MeV and 2.2 MeV which aligns remarkably well with the range necessary to address the $^7$Li problem (see Section \ref{sec:7Bedestruction}).

 \subsection{The confined $SU_c(2)$ phase and its vacuum}
 \label{appendix:confined-SUC(2)}
 
At temperatures below $T_{\rm LGB, m}$ is the $SU_c(2)$ phase which effectively constitutes a partial confinement within the 2SC phase . The deconfined-to-confined $SU_c(2)$ phase transition, as the temperature decreases within the 2SC phase, is second order (i.e. symmetry breaking has occurred; see \cite{ouyed_2001} and references therein). The 2SC without LGB decay belongs to a different global symmetry while after LGB decay one is left with the confined $SU_c(2)$ which breaks the $\mathbb{Z}_2$-symmetry  (see e.g. \cite{satz_2001} for phase transitions in QCD). While the flavour symmetries and other (e.g. Poincar\'e) symmetries remain the same during the transition from deconfined-to-confined $SU_c(2)$, the vacuum looses the $\mathbb{Z}_2$-symmetry associated with the center of the $SU_c(2)$  group and ``chooses” a specific direction or configuration. The vacuum of the confined $SU_c(2)$ is in principle different from that of the unconfined $SU_c(2)$  and it is not unreasonable to assume that it may be metastable. 

$\mathbb{Z}_2$-symmetry breaking and false vacuum metastability and tunnelling into the true vacuum with consequences to cosmology have been discussed in the literature (e.g. \cite{addazi_2019}  and references therein).  In our case, the phenomenon of tunnelling or leaking could be attributed to two potential factors:  (i)  the zero-momentum loop at the origin of zero-point energy at zero temperature; (ii) the gluon condensate if we allow for a small temperature that is insignificant compared to any other scale in the problem. Which mechanism is exactly at play is currently uncertain within the scope of our model and what is presented here serves as preliminary outlines for potential future research directions. The main point is that the leakage suggests a mechanism to explain $\Lambda$-CDM cosmology in our model.

\section{Our cosmology}
\label{appendix:our-cosmology}

Let us write again the  time evolution of the density  inside a SQC as given in Eq. (\ref{eq:rhosqc-evolution}):
 \begin{align}
 \label{appendix:rhosqc-evolution}
  \rho_{\rm sqc}(t)&=(1-\eta_{\rm G})\rho_{\rm csc}-\rho_{\rm QCD}^{\rm vac.}(1-e^{-(t-t_{\rm G})/t_{\rm tun.}})\\\nonumber
  &= \rho_{\rm sqc, eG} \left(1- \frac{f_{\rm V}}{1-\eta_{\rm G}}(1-e^{-(t-t_{\rm G})/t_{\rm tun.}})\right)\ ,
 \end{align}
 with $\rho_{\rm sqc, eG}  = (1-\eta_{\rm G}) \rho_{\rm csc}$  the SQC density at the end of the LGB/photon-burst phase.
  As noted earlier,  the equation above incorporates the key parameters in our cosmology namely, $\rho_{\rm csc}, \eta_{\rm G}, f_{\rm V}$ and $t_{\rm tun.}$ which are all fundamentally related to QCD.
 
 The time evolution of the DM density given in Eq. (\ref{eq:rhoDM})  can be expanded as

 \begin{equation}
\label{appendix:rhoDM}
\rho_{\rm DM}(t)  =
\begin{cases}
 \frac{\rho_{\rm sqc}(t) V_{\rm sqc}^{\rm tot.}}{V_{\rm univ.}(t)} = \frac{\rho_{\rm sqc}(t)}{\rho_{\rm sqc, 0}}\times \frac{\rho_{\rm sqc, 0} V_{\rm sqc}^{\rm tot.}}{V_{\rm univ., 0}}\times  \frac{V_{\rm univ., 0}}{V_{\rm univ.}(t)}\ & \\
 \quad\qquad\qquad\qquad\qquad {\rm if} \quad t > t_{\rm G}\ ({\rm or}\ z < z_{\rm G}) &\\
  \frac{\rho_{\rm csc} V_{\rm sqc}^{\rm tot.}}{V_{\rm univ.}(t)} = \frac{\rho_{\rm csc}}{\rho_{\rm sqc, 0}}\times \frac{\rho_{\rm sqc, 0} V_{\rm sqc}^{\rm tot.}}{V_{\rm univ., 0}}\times  \frac{V_{\rm univ., 0}}{V_{\rm univ.}(t)}\ & \\
  \quad\qquad\qquad\qquad\qquad {\rm if} \quad t \le t_{\rm G}\ ({\rm or}\ z \ge z_{\rm G})\\
 \end{cases}   
\end{equation}
where the subscript ``0" refers to the current age of the universe at $z=0$ (i.e. at $t=t_0$); recall that $t_{\rm tun.}$ is a fraction
of $t_0$ which simplifies our equations. $V_{\rm sqc}^{\rm tot.}$ is the total volume occupied by SQCs,
which is constant in time, and $V_{\rm univ.}(t)$ the Hubble volume at time $t$. 

  Putting  Eq. (\ref{appendix:rhosqc-evolution}) into Eq. (\ref{appendix:rhoDM}) we get  
  \begin{equation}
\label{appendix:rhoDM2}
\rho_{\rm DM}(t)  =
\begin{cases}
\rho_{\rm DM, 0}\frac{V_{\rm univ., 0}}{V_{\rm univ.}(t)}\times \frac{\left(1- \frac{f_{\rm V}}{1-\eta_{\rm G}}\left(1-e^{-(t-t_{\rm G})/t_{\rm tun.}}\right)\right)}{\left(1- \frac{f_{\rm V}}{1-\eta_{\rm G}}\right)} & \\
\quad\qquad\qquad\qquad\qquad   {\rm if} \quad t > t_{\rm G}\ ({\rm or}\ z < z_{\rm G}) & \\
  \rho_{\rm DM, 0}\frac{V_{\rm univ., 0}}{V_{\rm univ.}(t)}\times \frac{1}{(1-\eta_{\rm G}) -f_{\rm V}}\ & \\
 \quad\qquad\qquad\qquad\qquad {\rm if} \quad t \le t_{\rm G}\ ({\rm or}\ z \ge z_{\rm G})\\
 \end{cases}   
\end{equation}
with $\rho_{\rm sqc, 0}= (1-\eta_{\rm G})\rho_{\rm csc}-\rho_{\rm QVD}^{\rm vac.}$, $\rho_{\rm sqc, eG}/\rho_{\rm sqc, 0} = (1- \frac{f_{\rm V}}{1-\eta_{\rm G}})^{-1}$ and  $\rho_{\rm csc}/\rho_{\rm sqc, 0}=((1-\eta_{\rm G})-f_{\rm V})^{-1}$; see Section \ref{sec:SQCs-DE}.  Also, $\rho_{\rm DM, 0}= \rho_{\rm sqc, 0} V_{\rm sqc}^{\rm tot.}/V_{\rm univ., 0}$ is the DM density at $z=0$.  The jump in the value of $\rho_{\rm DM}$ at $t=t_{\rm G}$ in the equation above is due to the fact that a fraction $(1-\eta_{\rm G})$ of the DM is lost to the 2 MeV radiation 
before the start of the leakage era at $t_{\rm eG}$ (see Section \ref{sec:7Bedestruction}). 

Eq. (\ref{appendix:rhoDM2}) can be expressed in terms of redshift as
\begin{equation}
\label{appendix:rhoDM3}
\rho_{\rm DM}(z, z_{\rm tun.})  =
\begin{cases}
\rho_{\rm DM, 0}(1+z)^3\times \frac{\left(1- \frac{f_{\rm V}}{1-\eta_{\rm G}}\left(1-e^{-\frac{(1+z_{\rm tun.})^{3/2}}{(1+z)^{3/2}}}\right)\right)}{\left(1- \frac{f_{\rm V}}{1-\eta_{\rm G}}\right)} & \\
\quad\qquad\qquad\qquad\qquad  {\rm if} \quad z < z_{\rm G}\sim z_{\rm eq.} & \\
  \rho_{\rm DM, 0}(1+z)^3\times \frac{1}{(1-\eta_{\rm G}) -f_{\rm V}}\ & \\
\quad\qquad\qquad\qquad\qquad    {\rm if} \quad z \ge z_{\rm G}\sim z_{\rm eq.} &\\
 \end{cases}   
\end{equation}
where we emphasize the dependency of $\rho_{\rm DM}$ on the leakage characteristic redshift $z_{\rm tun.}$ to help 
differentiate our model from the $\Lambda$CDM cosmology.
Because $t_{\rm G} << t_{\rm tun.}$ (or $z_{\rm G} >> z_{\rm tun.}$) and most of the time between $t_{\rm G}$ and $t_{\rm tun.}$ is in the matter dominated era, we approximate  $t/t_{\rm tun.}\sim (1+z_{\rm tun.})^{3/2}/(1+z)^{3/2}$ and $V_{\rm univ., 0}/V_{\rm univ.}(t)\sim (1+z)^3$. 
 Our model does not depend critically on $z_{\rm G}$ and by setting $z_{\rm G}\sim z_{\rm eq.}$,  where 
   $z_{\rm eq.}$ is the redshift at matter-radiation equality,  we can simplify the equations without changing the final results. 

 The redshift of matter-radiation equality in our model is estimated by writing $\rho_{\rm DM}(z_{\rm eq.},z_{\rm tun.}) = \rho_{\rm r, 0}(1+z_{\rm eq.})^4$,
with   $t<< t_{\rm tun.}$ (i.e. $z>> z_{\rm tun.}$)
 so that $\rho_{\rm DM}(z_{\rm eq.}, z_{\rm tun.})\simeq \rho_{\rm DM, 0}(1+z_{\rm eq.})^3\times (1- \frac{f_{\rm V}}{1-\eta_{\rm G}})^{-1}$. This
yields
\begin{align}
1+z_{\rm eq.}&\sim \frac{\rho_{\rm DM, 0}}{\rho_{\rm r, 0}}\times \left(1- \frac{f_{\rm V}}{1-\eta_{\rm G}}\right)^{-1}\\\nonumber
&= \frac{\Omega_{\rm DM, 0}}{\Omega_{\rm r, 0}}\times \left(1- \frac{f_{\rm V}}{1-\eta_{\rm G}}\right)^{-1}\\\nonumber
 &=\frac{\omega_{\rm DM}}{\omega_{\rm r}}\times \left(1- \frac{f_{\rm V}}{1-\eta_{\rm G}}\right)^{-1}\ ,
\end{align}
with $\omega_{\rm DM}=\Omega_{\rm DM, 0} h^2\simeq 0.120$, $\omega_{\rm b}=\Omega_{\rm b, 0} h^2\simeq 0.0224$  and $\omega_{\rm r}=\Omega_{\rm r, 0} h^2\simeq 4.18\times 10^{-5}$ as measured by Planck \cite{planck_2018}; $h=H_0/100$  km s$^{-1}$ Mpc$^{-1}$ is the dimensionless Hubble constant in today's universe.
As expected, $z_{\rm eq.}$ in our model is larger (i.e. requires an increase in the radiation to compensate for larger DM in the past) than in the case of the pure  $\Lambda$CDM cosmology which does not capture the converted (by LGB decay and vacuum leakage) component of the DM.

Finally, our cosmology can be described by
\begin{align}
\label{Hzprepost}
H&(z, z_{\rm tun.}) = H_0\times\\\nonumber
&\times  \sqrt{\Omega_{\rm r, 0}(1+z)^4+ \Omega_{\rm b, 0}(1+z)^3 + \Omega_{\rm DM}(z, z_{\rm tun.})+\Omega_{\Lambda}}
\end{align}
with $\Omega_{\rm DM}(z, z_{\rm tun.})=\rho_{\rm DM}(z, z_{\rm tun.})/\rho_{\rm c}^0$ and $H_0^2=8\pi G\rho_{\rm c}^0/3$ where  $\rho_{\rm c}^0$ is today's critical density; 
$\Omega_{\Lambda}=1-\Omega_{\rm r, 0}-\Omega_{\rm b, 0}-\Omega_{\rm DM, 0}$.

The co-moving sound horizon and the co-moving angular-diameter distance to the  surface of last scatter are \cite{ryden_2016}
\begin{equation}
r_{\rm s}(z_{\rm tun.})=\int_{z_{ls}}^\infty \frac{c_{\rm s}(z) dz}{H(z, z_{\rm tun.})};\quad D_{\rm A}(z_{\rm tun.})= \int_0^{z_{ls}} \frac{c dz}{H(z, z_{\rm tun.})}\ .
\end{equation}
The sound speed of the photon-baryon fluid at $z\ge z_{ls}$ is $c_{\rm s}(z)  = c/\sqrt{3(1+R(z))}$ with
$R(z) = (3/4)(\omega_{\rm b}/\omega_\gamma)/(1+z)$ and $\omega_{\gamma}=2.47\times 10^{-5}$.

The Hubble constant $H_0$ is found by solving $\theta_{\rm s}D_{\rm A}(z_{\rm tun.})- r_{\rm s}(z_{\rm tun.})=0$
with $\theta_{\rm s}=1.041\times 10^{-2}$ the angle subtended by the sound horizon \cite{planck_2018}. For the $\Lambda$CDM one finds
$H_0\sim 67.3$ km s$^{-1}$ Mpc$^{-1}$ ($h\sim 0.673$; by solving $\theta_{\rm s}D_{\rm A}- r_{\rm s}=0$ using standard $\Lambda$CDM cosmology).
 In our case, $H_0\sim73$ km s$^{-1}$ Mpc$^{-1}$ ($h=0.73$) can be obtained for a range in $\eta_{\rm G}$ and $f_{\rm V}$ values (see
 examples in Figure \ref{fig:H0}) with a 
  leakage characteristic redshift  $2 < z_{\rm tun.} < 6$ (i.e. $1 < t_{\rm tun.} ({\rm Gyr})< 4$); see Section \ref{sec:SQCs-DE}.



\reftitle{References}


\begin{thebibliography}{99}


\bibitem[Hoyle \& Tayler(1964)]{hoyle_1964} Hoyle, F. \& Tayler, R. J.\ {\bf 1964},  {\em The Mystery of the Cosmic Helium Abundance}, {\em Nature} 203, 1108 


\bibitem[Peebles(1966)]{peebles_1966} Peebles, J. P. E.\ {\bf 1966}, {\em Primordial Helium Abundance and the Primordial Fireball. II}, {\em ApJ},  146, 542 


\bibitem[Wagoner et al.(1967)]{wagoner_1967} Wagoner, R. V., Fowler, W. A. \& Hoyle, F. {\bf 1967}, {\em On the Synthesis of Elements at Very High Temperatures}, {\em ApJ}, 148, 3 


\bibitem[Tytler et al.(2000)]{tytler_2000} Tytler, D., O'Meara, J. M., Suzuki, N. \& Lubin, D.\ {\bf 2000}, {\em Review of Big Bang Nucleosynthesis and Primordial Abundances}, {\em Phys. Scripta} T85, 12


\bibitem[Lang(1999)]{lang_1999} Lang, K.R., {\em Astrophysical Formulae},  Volume II: Space, Time, Matter and Cosmology,
Springer-Verlag, Berlin {\bf 1999}


\bibitem[Khatri \& Sunyaev(2011)]{khatri_2011} Khatri, R. \& Sunyaev, R.~A.\ {\bf 2011}, {\em Time of primordial $^7$Be conversion into $^7$Li, energy release and doublet of narrow cosmological neutrino lines}, {\em Astronomy Letters}, 37, 367


\bibitem[Spite \& Spite(1982)]{spite_1982} Spite, F. \& Spite, M.\ {\bf 1982}, {\em Abundance of lithium in un-evolved stars and old disk stars : Interpretation and consequences.}, {\em A\&A},115, 357


\bibitem[Fields(2011)]{fields_2011}  Fields, B. D.\ {\bf 2011}, {\em The Primordial Lithium Problem}, {\em Annual Review of Nuclear and Particle Science}, 61, 47 


\bibitem[Berezinsky et al.(1990)]{berezinsky_1990} Berezinsky V. S., Bulanov S. V., Dogiel V. A., Ginzburg V. L. \& Ptuskin V. S., {\bf 1990}, {\em Astrophysics of Cosmic Rays}, Amsterdam, Netherlands: North-Holland (1990) 534 p


\bibitem[Kawasaki \& Moroi(1995)]{kawasaki_1995} Kawasaki, M. \& Moroi, T.\ {\bf 1995}, {\em Electromagnetic Cascade in the Early Universe and Its Application to the Big Bang Nucleosynthesis}, {\em ApJ}, 452 506


\bibitem[Poulin \& Serpico(2015)]{poulin_2015} Poulin, V. \& Serpico, P. D.\ {\bf 2015}, {\em Loophole to the universal
photon spectrum in electromagnetic cascades: application to the ``cosmological lithium problem"}, {\em Phys. Rev. Lett.} 114, 091101 (PS15)


\bibitem[Protheroe et al.(1995)]{protheroe_1995} Protheroe, R.J., Stanev, T. \& Berezinsky, V.S.\ {\bf 1995}, {\em Electromagnetic cascades and cascade nucleosynthesis in the early Universe}, {\em Phys. Rev. D} , 51(8), 4134


 \bibitem[Kawasaki et al.(2020)]{kawasaki_2020}   Kawasaki, M., Kohri, K., Moroi, T., Murai, K., Murayama, H. \ {\bf 2020}, {\em Big-bang nucleosynthesis with sub-GeV massive decaying particles}, {\em JCAP}, 2020, 048.  doi:10.1088/1475-7516/2020/12/048
 

\bibitem[Witten(1984)]{witten_1984} Witten, E.\ {\bf 1984}, {\em Cosmic separation of phases}, {\em Phys. Rev. D}  30, 272


\bibitem[Zhitnitsky(2003)]{zhitnitsky_2003}  Zhitnitsky, A. R.\ {\bf 2003}, {\em `Nonbaryonic' dark matter as baryonic colour superconductor}, {\em JCAP}, 10, 10


\bibitem[Ishida et al.(2014)]{ishida_2014} Ishida, H., Kusakabe, M. \& Okada, H.\ {\bf 2014}, {\em Effects of long-lived 10 MeV-scale sterile neutrinos on primordial elemental abundances and the effective neutrino number}, {\em Phys. Rev. D}  90, 8, 083519


\bibitem[R{\"u}ster et al.(2004)]{ruster_2004} R{\"u}ster, S. B., Shovkovy, I. A. \& Rischke, D. H.\ {\bf 2004}, {\em Phase diagram of dense neutral three-flavor quark matter}, {\em Nucl. Phys. A} 743, 127


\bibitem[Alford et al.(2008)]{alford_2008} Alford, M. G., Schmitt, A., Rajagopal, K. \& Sch{\"a}fer, T.\ {\bf 2008}, {\em Color superconductivity in dense quark matter},  {\em Reviews of Modern Physics}, 80, 1455


\bibitem[Baym et al.(2018)]{baym_2018} Baym, G., Hatsuda, T.,  Kojo, T., Powell, P.D., Song, Y., Takatsuka, T.\ {\bf 2018},  {\em From hadrons to quarks in neutron stars: a review}, {\em Reports on Progress in Physics} 81, 056902


\bibitem[Brodsky et al.(2012)]{brodsky_2012} Brodsky, S.~J., Roberts, C. D., Shrock, R. \& Tandy, P. C.\ {\bf 2012},  {\em Confinement contains condensates}, {\em Phys. Rev. C}  85, 065202


\bibitem[Coleman(1977)]{coleman_1977} Coleman, S. R.\ {\bf 1977}, {\em Fate of the false vacuum: Semiclassical theory}, {\em Phys. Rev. D}  15, 2929; Erratum, {\em Phys. Rev. D}  16, 1248(E) (1977)


\bibitem[Cohen-Tannoudji et al.(2006)]{tannoudji_2006} Cohen-Tannoudji, C., Diu, B. \& Laloe, F. \ {\bf 2006}, {\em Quantum Mechanics}. (Hermann and John Wiley \& Sons, Paris, France)


\bibitem[Kamionkowski \& Riess(2022)]{kamionkowski_2022}  Kamionkowski, M \& Riess, A. G.\ {\bf 2023}, {\em The Hubble Tension and Early Dark Energy}, {\em Ann. Rev. Nucl. Part. Sci.}  73, 153


\bibitem[Sch{\"o}neberg et al.(2022)]{schoneberg_2022} Sch{\"o}neberg, N.,  Abell{\'a}n, G. F., S{\'a}nchez, A. P.,  Witte, S. J., Poulin, V. \& Lesgourgues, J.\ {\bf 2022},  {\em The $H_0$ 
 Olympics: A fair ranking of proposed models},  {\em Physics Reports} 984, 1


\bibitem[Garrett \& Gintaras(2011)]{garrett_2011}   Garrett, K. \& Gintaras, D.\ {\bf 2011}, {\em Dark Matter: A Primer}, {\em Advances in Astronomy}, 2011, 968283


\bibitem[Schumann(2015)]{schumann_2015}  Schumann, M.\ {\bf 2015}, {\em Dark Matter 2014}, {\em European Physical Journal Web of Conferences}, 96, 01027


\bibitem[Bertone \& Hooper(2018)]{bertone_2018} Bertone, G. \& Hooper, D.\ {\bf 2018}, {\em History of dark matter}, {\em Reviews of Modern Physics} 90, 045002


\bibitem[Kisslinger \& Das(2019)]{kisslinger_2019}  Kisslinger, L. S. \& Das, D.\ {\bf 2019}, {\em A brief review of dark matter}, {\em International Journal of Modern Physics} A, 34, 1930013


\bibitem[Oks(2021)]{oks_2021}  Oks, E.\ {\bf 2021}, {\em Brief Review of Recent Advances in Understanding Dark Matter and Dark Energy}, {\em New Astronomy Reviews}, 93, 101632


\bibitem[Arbey \& Mahmoudi(2021)]{arbey_2021} Arbey, A. \& Mahmoudi, F.\ {\bf 2021}, {\em Dark matter and the early Universe: A review}, {\em Progress in Particle and Nuclear Physics}, 119, 103865


\bibitem[Super-Kamiokande Collaboration et al.(2002)]{superK_2002} Super-Kamiokande Collaboration et al.\  {\bf 2002}, {\em Search for proton decay via $p\rightarrow \mu^+K^0$ in 0.37 megaton-years exposure of Super-Kamiokande} [arXiv:2208.13188]


\bibitem[Sofue(2020)]{sofue_2020} Sofue, Y.\ {\bf 2020}, {\em Rotation Curve of the Milky Way and the Dark Matter Density}, Galaxies 8, Issue 2, id.37


\bibitem[Kanno\&Soda(2012)]{kanno_2012}  Kanno, S. \& Soda, J.\ {\bf 2012}, {\em {\em Exact Coleman-De Luccia instantons},}, {\em Int. J. Mod. Phys.} D 21, 1250040

 
 \bibitem[Navarro, Frenk \& White(1996)]{navarro_1996} Navarro J. F., Frenk C. S., White S. D. M.\ {\bf 1996}, {\em The Structure of Cold Dark Matter Halos}, {\em ApJ}, 462, 563


\bibitem[Abel et al.(2002)]{abel_2002} Abel T., Bryan G. L., Norman M. L., {\bf 2002}, {\em The Formation of the First Star in the Universe}, {\em Science}, 295, 93


\bibitem[Paczy\'nski(1972)]{paczynski_1972} Paczy\'nski, B.\ {\bf 1972}, {\em Carbon Ignition in Degenerate Stellar Cores}, {\em ApJ} L11, 53


\bibitem[Ablyazimov et al.(2017)]{ablyazimov_2017}  Ablyazimov T et al.\ {\bf 2017}, {\em Challenges in QCD matter physics --The scientific programme of the Compressed Baryonic Matter experiment at FAIR}, {\em Eur. Phys. J.} A 53 60


\bibitem[Senger(2020)]{senger_2020} Senger, P.\ {\bf 2020}, {\em Status of the Compressed Baryonic Matter experiment at FAIR}, {\em International Journal of Modern Physics} E, 29, 2030001


\bibitem[Huang \& Shovkovy(2004)]{huang_2004} Huang, M. \& Shovkovy, I. A.\ {\bf 2004}, {\em Chromomagnetic instability in dense quark matter}, {\em Phys. Rev. D}  70, 051501(R); Phys. Rev. D 70, 051501


\bibitem[Alford et al.(2005)]{alford_2005}    Alford, M., Kouvaris, C. \& Rajagopal, K.\ {\bf 2005},   {\em Evaluating the gapless color-flavor locked phase}, {\em Phys. Rev. D} 71, 054009


\bibitem[Cao et al.(2015)]{cao_2015}  Cao, G., He, L. \&  Zhuang, P.\ {\bf 2015},  {\em Solid-state calculation of crystalline color superconductivity}, {\rm Phys. Rev.} D 91, 114021


\bibitem[Haber \& Schmitt(2018)]{haber_2018} Haber, A. \& Schmitt, A. \ {\bf 2018}, {\em New color-magnetic defects in dense quark matter}, {\em Journal of Physics G Nuclear Physics} 45, 065001


\bibitem[Guenther(2021)]{guenther_2021} Guenther, J. N.\ {\bf 2021}, {\em Overview of the QCD phase diagram -- Recent progress from the lattice}, {\em Eur. Phys. J. A} 57, 136


\bibitem[Alford(2001)]{alford_2001b} Alford, M.\ {\bf 2001},  {\em Color superconducting quark matter}, {\em Annu. Rev. Nucl. Part. Sci.} 51, 131


\bibitem[Svensson \& Zdziarski(1990)]{svensson_1990} Svensson, R. \& Zdziarski, A. A.\ {\bf 1990}, {\em Photon-Photon Scattering of Gamma Rays at Cosmological Distances}, {\em ApJ}, 349, 415


\bibitem[Ouyed \& Sannino(2001)]{ouyed_2001} Ouyed, R., \& Sannino, F. {\bf 2001}, {\em The Glueball sector of two-flavor Color Superconductivity
}, {\em Phys. Lett. B. 511, 66}


\bibitem[Sannino et al.(2002)]{sannino_2002} Sannino, F., Marchal, N. \& Sch\"afer, W. {\bf 2002},  {\em Partial Deconfinement in Color Superconductivity}, {\em Phys. Rev. D}  66, 016007


\bibitem[Rischke et al.(2001)]{rischke_2001} Rischke, D. H. Son, D. T. \&. Stephanov, M. A.\ {\bf 2001}, {\em Asymptotic Deconfinement in High-Density QCD}, {\em Phys. Rev. Lett.}  87, 062001


\bibitem[Alford et al.(2001)]{alford_2001} Alford, M.,G. Rajagopal, K., Reddy, S. \& F. Wilczek, F.\ {\bf 2001}, {\em Minimal color-flavor-locked–nuclear interface}, {\em Phys. Rev. D} 64, 074017 


\bibitem[Shovkovy(2004)]{shovkovy_2004} Shovkovy, I. A. {\bf 2004}, {\em lectures delivered at the IARD 2004 conference, Saas Fee, Switzerland}, June 12-19, and at the Helmholtz International Summer School and Workshop on Hot points in Astrophysics and Cosmology, JINR, Dubna, Russia, Aug. 2-13 [arXiv:nucl-th/0410091]


\bibitem[Satz(2001)]{satz_2001} Satz, H.\ {\bf 2001}, {\em Phase transitions in QCD}, {\em Nucl. Phys.} A 681, 3


\bibitem[Addazi, et al.(2019)]{addazi_2019} Addazi, A., Marcian{`o}, A., Pasechnik, R. \& Prokhorov, G.\ {\bf 2019}, {\em Mirror symmetry of quantum Yang-Mills vacua and cosmological implications}, {\em European Physical Journal} C 79, 251


\bibitem[Planck collaboration(2018)]{planck_2018} Planck collaboration\ {\bf 2018}, {\em Planck 2018 results. VI. Cosmological parameters}, {\em A\&A} 641, A6 


\bibitem[Ryden(2016)]{ryden_2016} Ryden, B.\ {\bf 2016}, {\em Introduction to Cosmology},  2nd Edition (Cambridge University Press, Cambridge)


\end{thebibliography}
\end{document}